\documentclass[12pt,psfig,epsfig,graphicx,psfrag]{article}
\usepackage{graphicx}
\usepackage{amsfonts,amssymb,amsmath}
\usepackage{psfrag}

\newcommand{\be}{\begin{equation}}
\newcommand{\ee}{\end{equation}}
\newcommand{\ba}{\begin{eqnarray}}
\newcommand{\ea}{\end{eqnarray}}

\newcommand{\arctanh}{\mathrm{arctanh}}
\def\L5{\tilde{\Lambda}}

\renewcommand{\d}{{\mathrm{d}}}

\def\1{\mathchoice{\rm 1\mskip-4.2mu l}{\rm 1\mskip-4.2mu l}%
{\rm 1\mskip-4.6mu l}{\rm 1\mskip-5.2mu l}}

\begin{document}

\begin{flushright}
\end{flushright}
\vskip 1cm

\begin{center}
{\Large {\bf Causal structure of bigravity
solutions}}\\[1cm]
D.~Blas$^c$\footnote{dblas@ffn.ub.es},
C.~Deffayet$^{a,b,}$\footnote{deffayet@iap.fr},
J.~Garriga$^{c,}$\footnote{garriga@ifae.es},
\\
$^a${\it APC\;\footnote{UMR 7164 (CNRS, Universit\'e Paris 7, CEA, Observatoire de Paris)}, 11 place Marcelin Berthelot,\\
 75005 Paris Cedex 05, France.}\\
 $^b$ {\it GReCO/IAP\;\footnote{UMR 7095 (CNRS, Universit\'e Paris 6)}
 98 bis boulevard Arago, 75014 Paris, France.}\\
$^c${\it Departament de F\'isica Fonamental, Universitat de Barcelona,\\
Diagonal 647, 08028 Barcelona, Spain.}
\\
\end{center}
\vskip 0.2cm

\noindent
\begin{abstract}

We discuss the causal diagrams of static and spherically symmetric
bigravity vacuum solutions, with interacting metrics $f$ and $g$.
Such solutions can be classified into type I (or "non-diagonal")
and type II (or "diagonal"). The general solution of type I is
known, and leads to metrics $f$ and $g$ in the
Schwarzschild-(Anti)de Sitter family. The two metrics are not
always diagonalizable in the same coordinate system, and the
light-cone structure of both metrics can be quite different. In
spite of this, we find that causality is preserved, in the sense
that closed time-like curves cannot be pieced together from
geodesics of both metrics. We propose maximal extensions of Type I
bigravity solutions, where geodesics of both metrics do not stop
unless a curvature singularity is encountered. Such maximal
extensions can contain several copies (or even an infinite number
of them) of the maximally extended "individual" geometries
associated to $f$ and $g$ separately. Generically, we find that
the maximal extensions of bigravity solutions are not globally
hyperbolic, even in cases when the individual geometries are. The
general solution of type II has not been given in closed form. We
discuss a subclass where $g$ is an arbitrary solution of
Einstein's equations with a cosmological constant, and we find
that in this case the only solutions are such that $f\propto g$
(with trivial causal structure).

\end{abstract}

\pagebreak

\section{Introduction}

 Recently, massive gravity, and other infrared
modifications of General Relativity, have been the subject of
numerous
investigations\cite{Dvali:2000hr,Damour:2002ws,Arkani-Hamed:2003uy,
Rubakov:2004eb,Dubovsky:2004sg}, largely motivated by the enigma
of dark energy\cite{ACCDGP,Damour:2002wu,Dubovsky:2005dw}. A
linearized ghost-free theory of massive gravity can be constructed
by adding a suitable quadratic mass term to the linearized
Einstein Lagrangian. This is the so-called Pauli-Fierz theory (PF)
\cite{Fierz:1939ix}. Non-linear and generally covariant extensions
can easily be formulated, at the cost of introducing new fields.
The simplest possibility is to consider the theory of two metrics,
$f$ and $g$, with standard kinetic terms, where the mass terms
arise from non-derivative interactions between $f$ and $g$. This
is known as bigravity, $f-g$ theory, or {\it strong gravity}, and
was first considered in the seventies, in the context of the
strong interactions \cite{Isham:gm}\footnote{Bigravity (and
multigravity in general) is also relevant to the program of
``deconstruction'' of gravity \cite{DECONS,GRAVDECONS}}. Massive
gravity poses certain difficulties. First of all, the linearized
theory does not agree with observations. The PF Lagrangian
propagates a scalar degree of freedom (in addition to two tensor
and two vector polarizations) which couples to matter with
gravitational strength. The coupling persists in the limit as the
graviton mass vanishes, causing the so-called vDVZ discontinuity
\cite{vDVZ}. It has been suggested that the discontinuity may
disappear due to non-linear effects \cite{Vainshtein:1972sx}, in
the sense that solutions of massless gravity can be recovered
non-perturbatively. This mechanism may indeed work in more
complicated infrared modifications of gravity, such as the
theories discussed in \cite{CUREVDVZ}, although its implementation
in the context of bigravity may be problematic (see e.g.
\cite{Damour:2002gp}). In addition, the minimal non linear
extension of PF theory was found to propagate six degrees of
freedom rather than five \cite{Boulware:my}, the sixth degree of
freedom being ghost like. This increase in the number of
propagating degrees of freedom was also found to occur in
bigravity \cite{Damour:2002ws},  and in general, it may be related
to the strong self-interaction of the scalar polarization of the
graviton \cite{Deffayet:2005ys,Creminelli:2005qk}.
%sixth one is ghost-like. Nevertheless, it has been argued that
%massive gravity makes sense as an effective theory below a certain
%cut-off scale \cite{nima}.

Despite such difficulties, bigravity remains an elegant framework
where to investigate the massive graviton, particularly because a
large class of exact solutions to this theory are known.
Interestingly, there are solutions where both metrics $f$ and $g$
belong to the Schwarzschild-(Anti) de Siter family, which makes
them phenomenologically attractive. On the other hand, in dealing
with a space-time with two metrics, it is natural to ask whether
we can make sense of its causal structure. In general, the
light-cones generated by metrics $f$ and $g$ will not agree, and
this may lead to pathologies which may restrict the class of
physically acceptable solutions. The purpose of this paper is to
investigate this issue in some detail.

In Section 2, we review the classification of static spherically
symmetric solutions of bigravity. These fall into two types. In
type I solutions, the area of orbits of spherical symmetry as
measured by the $f$ metric is $2/3$  of the
area as measured by the $g$ metric. Otherwise, the solutions are
called type II. In type II solutions, both metrics can be brought
to diagonal form in the same coordinate system, whereas this is
not always possible in type I solutions, where only one of them
can be diagonalized at a time. The general solution of type I was
given in early work
\cite{Salam:1976as,Isham:1977rj,Damour:2002gp}.  The general
solution of type II is not known explicitly \cite{arache}. We
shall consider the subclass where one of the metrics is a solution
of Einstein's equations with a cosmological constant, for which we
discuss the general solution. In Section 3, we discuss the causal
structure of type I solutions. Section 4 summarizes our
conclusions. Some technical issues concerning type II solutions
are left to an Appendix.

\section{Spherically symmetric solutions in bigravity}

As mentioned in the introduction all type I solutions are known
analytically, while the general type II solution is not known. In
this Section, we review this classification, and comment on a
subset of type II solutions.

Following \cite{Isham:gm}, we consider the action
\begin{equation}
\label{action} S=\int \d^4 x\sqrt{-g}\ \left(\frac{- R_g}{2
\kappa_g} +L_g\right) +\int \d^4 x \sqrt{-f}\ \left(\frac{- R_f}{2
\kappa_f} + L_f\right) + S_{int}[f,g]
\end{equation}
Here $L_f$ and $L_g$ denote generic matter Lagrangians coupled to
the metrics $f$ and $g$ respectively. In solving the field
equations, we shall restrict attention to the case where there is
only a vacuum energy term in each matter sector $L_f=-\rho_f,
L_g=-\rho_g$. There is much freedom in the choice of the
interaction term in (\ref{action}). For definiteness, we shall
consider the form \cite{Isham:gm}
\begin{equation}\label{interaction}
 S_{int}=-\frac{\zeta}{4}\int
\d^4 x
(-g)^u(-f)^v(f^{\mu\nu}-g^{\mu\nu})(f^{\sigma\tau}-g^{\sigma\tau})(g_{\mu\sigma}
g_{\mu\tau}-g_{\mu\nu}g_{\sigma\tau})
\end{equation}
 with \be u+v = \frac{1}{2}.
\end{equation}
This reduces to the Pauli-Fierz one in the linear regime, when $g$
is taken to be the Minkowski metric (but many other choices could
be made with this same property \cite{Damour:2002ws}). The
equations of motion, derived from action (\ref{action}), read \be
G_{\mu \nu}^f = \kappa_f T_{\mu \nu}^f, \quad\quad G_{\mu \nu}^g =
\kappa_g T_{\mu \nu}^g, \label{e2} \ee where $G_{\mu\nu}$ denotes
the Einstein tensor, and the energy-momentum tensors are given by
\ba T_{\mu \nu}^f& = \frac{\zeta}{2} \left( \frac{g}{f}\right)^u
&\left[v f_{\mu \nu} (f^{\alpha \beta}-g^{\alpha \beta})
(f^{\sigma \tau} - g^{\sigma \tau})(g_{\alpha \sigma} g_{\beta
\tau}
- g_{\alpha \beta} g_{\sigma \tau}) \right.\nonumber\\
&&\left.- 2 (f^{\alpha \beta} - g^{\alpha \beta})(g_{\mu \alpha} g_{\beta \nu} -
g_{\mu \nu} g_{\alpha \beta})\right] + \rho_f f_{\mu\nu} \label{Tf}\\
T_{\mu \nu}^g &=  \frac{\zeta}{2} \nonumber \left(
\frac{f}{g}\right)^v &\left[ (f^{\alpha \beta} - g^{\alpha
\beta})(f^{\sigma \tau} - g^{\sigma \tau}) (u g_{\mu \nu}
g_{\alpha \sigma} g_{\beta \tau}- u g_{\mu \nu} g_{\alpha \beta}
g_{\sigma \tau} + 2 g_{\mu \alpha} g_{\sigma \nu} g_{\beta \tau} -
2 g_{\mu \alpha} g_{\nu \beta} g_{\sigma \tau})\right.\\&& \left.+
2 (f^{\alpha \beta} - g^{\alpha \beta}) (g_{\mu \alpha} g_{\beta
\nu} - g_{\mu \nu} g_{\alpha \beta})\right]+\rho_g g_{\mu\nu}
\label{Tga} \ea

Let us now concentrate on static spherically symmetric solutions.
By suitable choice of coordinates, the most general static and
spherically symmetric ansatz for bigravity can be cast into the
form
\ba
\label{formg} g_{\mu \nu} \d x^\mu \d x^\nu &=&
J \d t^2 - K \d r^2 - r^2 \left(\d \theta^2 + \sin ^2 \theta \; \d\phi^2 \right) \\
f_{\mu \nu} \d x^\mu \d x^\nu &=& C \d t^2 - 2 D \d t \d r - A \d r^2 -
B\left(\d \theta^2 + \sin ^2\theta \; \d \phi^2\right),
\label{formf} \ea where the metric coefficients are functions of
$r$. As noted in Ref. \cite{Isham:1977rj}, with the present ansatz
we have \be D G^f_{tt} + C G^f_{tr}=0. \ee Hence, from the
equations of motion, the following combination of components of
the energy momentum tensor $T^f$ must also vanish \be D T^f_{tt}+
CT^f_{tr}=-\frac{\zeta DJ}{2B} \left(\frac{JK r^4}{\Delta
B^2}\right)^u (3B -2 r^2) = 0. \ee Here, we have introduced
$\Delta = AC + D^2$ \cite{Isham:1977rj}. There are two different
ways of satisfying this equality. Either \be \label{typeI}
B=\frac{2}{3}r^2, \ee or \be D=0 \label{typeII}. \ee Solutions
obeying the first condition (\ref{typeI}) are called type I, and
they were first discussed in \cite{Salam:1976as}. The remaining
ones are called type II and, from (\ref{typeII}), they are
necessarily diagonal. Let us discuss both types of solutions in
turn.

\subsection{Type I solutions}

 In the case when the vacuum energies of matter fields vanish,
the general solution of type I was given in
\cite{Salam:1976as,Isham:1977rj}. The same method can be used in
the case when $\rho_f$ and $\rho_g$ are nonvanishing, and one
readily finds that the general solution is of the form \ba g_{\mu
\nu} \d x^\mu \d x^\nu &=& \left(1-q\right) \d t^2
- (1-q)^{-1} \d r^2 - r^2 (\d \theta^2 + \sin^2 \theta \d \phi^2) \label{spheg}\\
f_{\mu \nu} \d x^\mu \d x^\nu &=& \frac{2}{3\beta} (1-p) \d t^2 -2
D \d t \d r
 - A \d r^2  - \frac{2}{3} r^2 (\d\theta^2 + \sin^2 \theta \d\phi^2), \label{sphef}
\ea where
\ba
A &=& \frac{2}{3 \beta} (1-q)^{-2}\left(p + \beta - q- \beta q \right),\\
\label{Dequation} D^2 &=& \left(\frac{2}{3
\beta}\right)^2(1-q)^{-2}(p-q)(p+\beta-1 -\beta q). \ea Here
$\beta>0$ is an arbitrary constant, and the potentials $p$ and $q$
are functions of $r$. Substituting this ansatz into the expresions
(\ref{Tf}-\ref{Tga}), it can be checked that the energy momentum
tensors take the form of cosmological terms: $ T_{\mu\nu}^f =
 (\Lambda_f/\kappa_f) f_{\mu\nu}$, and $ T_{\mu\nu}^g =
 (\Lambda_g/\kappa_g) g_{\mu\nu}$, where \ba \frac{\Lambda_f}{\kappa_f} &=& \frac
{\zeta }{4} \left( \frac{3}{2}\right)^{4u}\beta^u \left\{3 v +
9\beta(1-v)\right\}
+ \rho_f,\label{lambafI} \\
\label{lambagI} \frac{\Lambda_g}{\kappa_g} &=& \frac
{\zeta}{4}\left( \frac{2}{3}\right)^{4v} \beta^{-v} \left\{ 3 u -
9 \beta (1+u)\right\}+\rho_g. \ea It is clear from the equations
of motion (\ref{e2}) that the metrics $f$ and $g$ must belong to
the Schwarzschild-(A)dS family. Note that the corresponding
cosmological constants (\ref{lambafI}-\ref{lambagI}) are not
determined solely by the vacuum energies $\rho_f$ and $\rho_g$.
They also contain a contribution from the interaction term in the
Lagrangian. This contribution depends not only on the parameters
$\zeta$ and $u$ (recall that $v=1/2 -u$), but also on the
arbitrary integration constant $\beta$.

 It is somewhat surprising that the cosmological
constants depend on an integration constant. This situation is
reminiscent of unimodular gravity, where the equations of motion
are the traceless part of Einstein's equations. In this case, the
vacuum energy does not determine the trace of the Ricci scalar
\cite{vanderBij} and Einstein's equations are recovered after
using of the Bianchi identities except for an arbitrary
integration constant that can chosen at will \cite{Zee:1985} (see
also \cite{Weinberg:1988cp}). One difference here is that we have
two cosmological constants $\Lambda_f$ and $\Lambda_g$, and we can
only choose the value of one of them at will.

% On the other hand, as we shall discuss at the beginning of
%Section III, Type I solutions tend to develop a specific type of
%singularity where one of the metrics becomes complex, unless we
%choose the particular value $\beta=1$. There are specific
%solutions for which this singularity is absent for $\beta\neq 1$,
%but they do not seem to be sufficiently generic. Therefore, it
%remains unclear at the moment whether the apparent freedom in
%shifting one of the $\Lambda$ will be of much practical use.

The metric (\ref{sphef}) can be put in a more familiar form
defining a new time coordinate $\tilde{t}$ by \be \label{ttilde}
\d\tilde{t} = \frac{1}{\sqrt{\beta}}\left \{\d t + \epsilon_D
\frac{\sqrt{(p-q)(p+\beta-1-\beta q)}}{(1-q)(1-p)}\d r \right\},
\ee  where $\epsilon_D = \pm 1 $ is defined by the sign
retained for $D$ from equation (\ref{sphef}), namely by \be D = -
\epsilon_D \frac{2}{3 \beta} (1-q)^{-1}\sqrt{(p-q)(p+ \beta - 1 -
\beta q)}. \ee  With such a coordinate change, the line
element (\ref{sphef}) now reads  \ba \label{fdiagonal} f_{\mu
\nu} \d x^\mu \d x^\nu = \frac{2}{3} \{(1-p) \d\tilde{t}^2 -
(1-p)^{-1} \d r^2 - r^2 (\d\theta^2 + \sin^2 \theta \d\phi^2)\}.
\ea As is clear from the previous discussion, the potentials $p$
and $q$ will be given by the familiar Schwarzschild-(A)dS forms
\ba
p&=& \frac{2 M_f}{r} + \frac{2 \Lambda_f}{9}r^2, \\
q &=& \frac{2 M_g}{r} + \frac{\Lambda_g}{3} r^2,
\ea where $M_f$ and
$M_g$ are two additional integration constants with the
interpretation of mass parameters.

It is tempting to conclude that this non-linear "theory of massive
gravity" is phenomenologically sound, since the vacuum solutions
of General Relativity with a cosmological term are recovered,
without a trace of the vDVZ discontinuity. In this sense, the mass
term does not seem to act as an exponential cut-off at a finite
range. Rather, it contributes to the effective cosmological
constant, which tends to bend space-time on a lengthscale of the
order of the inverse mass of the graviton (which is of order $m^2
\sim \kappa\ \zeta$)\footnote{{ See also the related discussion of
Ref. \cite{Gabadadze:2003jq}}}. On the other hand, this
contribution from the interaction term can be compensated for by a
finely-tuned contribution from the vacuum energy of matter fields,
and then we can have an asymptotically flat solution with exactly
the same form as for massless gravity.

It is therefore of some interest to understand the global
structure of such solutions, whose analysis we defer to the next
Section. The perturbative analysis and the investigation of
stability of these solutions are left for future work \cite{BDG}.
Let us now turn our attention to diagonal solutions.

\subsection{Type II solutions}

Type II solutions are defined as those which do not satisfy
(\ref{typeI}), and therefore must be diagonal. Although the
general type II solution is not known, some progress can be made
by further assuming that one of the metrics, say $g$, is a
solution of \be \label{Einstein} G_{\mu \nu}^g = \Lambda g_{\mu
\nu}, \ee for some constant $\Lambda$. In the Appendix, we show
that the only Type II solutions in this subclass are such that \be
\label{diag} f_{\mu\nu}=\gamma g_{\mu\nu}, \ee where $\gamma$ is a
constant (to be determined below).

Substituting (\ref{diag}) in (\ref{Tf}-\ref{Tga}) yields a
diagonal energy-momentum tensor for both the $g$ and the $f$
metrics, with respective cosmological constants given by \ba
\label{lambdag} \Lambda_g=-6 \kappa_g \zeta \gamma^{4v}
(-1+\gamma)(-1-2u+2
\gamma u)/(2\gamma^2)+\kappa_g \rho_g,\\
\label{lambdaf} \Lambda_f=-6 \kappa_f\zeta \gamma^{-4u}
(-1+\gamma)(1-2v+2 \gamma v)/(2\gamma^2)+\kappa_f \rho_f. \ea Eq.
(\ref{diag}) implies that the Einstein tensors for both metrics
are identical, and thus \be \Lambda_g=\gamma\Lambda_f.
\label{findg}\ee
This is an algebraic equation which determines
$\gamma$ in terms of the parameters in the Lagrangian.

For instance, in the case where the vacuum energies in both matter
Lagrangians are set to zero $\rho_g=\rho_f=0$, Eq. (\ref{findg})
becomes \be \label{gammaeq} (\gamma-1)(-2
u+(-1-\kappa+2u(1-\kappa))\gamma+2 \kappa u \gamma^2)=0. \ee  Here $\kappa = \kappa_g/\kappa_f$. In Ref.
\cite{arache}, the case with $u=0$ was considered. In that
particular case, the only solution of (\ref{gammaeq}) is
$\gamma=1$, and therefore $\Lambda_f=\Lambda_g=0$. This theory has
solutions with $f=g$, where $f$ is an arbitrary solution of the
vacuum Einstein equations \cite{arache}, and the maximally
symmetric background can only be flat space. However, for $u\neq
0$, Eq. (\ref{gammaeq}) has additional zeroes corresponding to the
vanishing of the second factor, and therefore we have additional
maximally symmetric bi-de Sitter or bi-AdS solutions (besides the
flat space solution $\gamma=1$), even in the case where the vacuum
energies $\rho_f$ and $\rho_g$ vanish.

Solutions satisfying (\ref{Einstein}) and (\ref{diag}) exist for
arbitrary interaction between both metrics [not necessarily of the
form (\ref{interaction})]. The reason is that by assuming the
ansatz where $f_{\mu\nu}=\gamma g_{\mu\nu}$, we necessarily have
$$T^{g}_{\mu\nu}={2\over \sqrt{g}}{\delta S_{int}[f,g]
\over \delta g^{\mu\nu}}=\Lambda_g(\gamma)\ g_{\mu\nu},$$ with
constant $\Lambda_g$, and similarly for $T^{f}$. All that will
change from one theory to the other is the form of the equation
$\Lambda_g(\gamma) = \gamma \Lambda_f(\gamma)$ which determines
$\gamma$. Maximally symmetric solutions of the form (\ref{diag})
in a class of bi-gravity theories have been previously considered
in \cite{Damour:2002wu}. A similar argument can be made for
multigravity theories, where one can always find solutions where
all metrics are proportional to each other (see e.g.
\cite{Deffayet:2004ws}).

As in the case of type I, both $f$ and $g$ solve Einstein's
equations with a cosmological term. In this sense, there is no
sign of a vDVZ discontinuity either, and one finds the usual
Schwarzschild-(A)dS metrics, without modification. Unlike Type II
solutions, here the two metrics share the same light cones and
therefore their causal structure doesn't pose any novelties. The
phenomenology and perturbative analysis of such solutions will be
discussed in a separate publication \cite{BDG}. \footnote{Assuming
that (\ref{Einstein}) is valid, these are the only solutions of
type II. This does not contradict the perturbative approach
developed by Aragone and Chela-Flores \cite{arache}, who
considered spherically symmetric and diagonal solutions for $f$
taking $g$ as a fixed flat metric. The difference with the present
case is that in their analysis $g$ is not dynamical, and therefore
there is no need for the tensor $T_g$ to be proportional to the
metric [as we have assumed, on account of (\ref{e2}) and
(\ref{Einstein})]. The analysis presented in the Appendix does not
exclude the existence of diagonal spherically symmetric bigravity
solutions, where $f$ would have an asymptotic behaviour along the
lines of \cite{arache}. It only shows that if such solutions
exist, then the metric $g$ will not be proportional to $f$; and
neither one will be a solution of Einstein's equations with a
cosmological constant.}

\section{Global structure of Type I solutions}

In dealing with spacetimes with two different metrics, it is
natural to worry about their compatibility from the causal point
of view. For instance, there is a Type I solution where $f$ is de
Sitter, while $g$ is Schwarzschild. The conformal diagrams of $f$
and $g$ look rather different, and the question is whether they
can be combined at all. The purpose of this section is to
investigate such combined causal diagrams, and the peculiarities
which they may introduce. We shall see, for instance, that when
the solution above is maximally extended, one can send a signal
from a geodesically complete de Sitter space to another
geodesically complete de Sitter space, by following geodesics of
the companion Schwarzschild metric. Another peculiarity is that
the maximal extensions we shall consider tend to raise issues of
global hyperbolicity, even in cases where both metrics $f$ and $g$
are separately globally hyperbolic. Nevertheless, we find that
there are no blatant violations of causality, in the sense that it
is not possible to construct closed future directed time-like
curves by combining geodesics of both metrics. In what follows, we
shall illustrate the construction of the diagrams through three
different examples.

 The basic procedure is to
maximally extend the geodesics of each metric, even when they
reach the conformal boundary of the companion metric. The
resulting causal structure will be illustrated by representing
light-cones of one of the metrics in the conformal diagram of the
other. The relevance of such light-cones is two-fold. First of
all, matter which is coupled to one of the metrics must follow
trajectories which are always inside of the future light cone from
a given point. Second, in the limit of very short wavelength, the
interaction term between both metrics becomes negligible compared
with the graviton kinetic term. Hence, high frequency gravitons
will follow null geodesics of the corresponding metric.

Before we proceed with the specific examples, we should briefly
comment on a particular type of "singularity" which arises in
certain bigravity solutions (again, even in cases where both
metrics are separately smooth). Note that the metric (\ref{sphef})
becomes complex in regions where $D^2<0$. As noted in
\cite{Isham:1977rj}, the coordinate singularity at $D=0$ can be
removed by a change of variables. This is of course true, since
$f$ is in the family of Schwarzschild-(A)dS metrics, which are
everywhere smooth (except perhaps at $r=0$ when $M_f \neq 0$).
 However, it does not seem to be possible to find a
change of variables which would remove the singularity from both
metrics at once, in the vicinity of the point at which $D^2$
changes sign, and which would make both metrics real. The reason
is that there are geodesics of $g$ which invade the regions
$D^2<0$ (with arbitrary slope, in fact). On such geodesics, the
line element with respect to $f$ is generically complex, and since
the line element is a scalar, this fact cannot be changed by a
coordinate transformation.
%by considering the simpler system, \ba
%\d s_1^2&=&\d t^2-\sqrt u \d u \d t- \d u^2,\\
%\d s_2^2&=&\d t^2- \d u^2. \ea Here, both metrics are flat, and
%therefore smooth when separately considered. However, there is no
%analytic change of variables such that: (a) $\det[g_1] \neq 0$ and
%$\det[g_2] \neq 0$ in an open set containing $u=0$, (b) Both
%metrics are well behaved in a neighborhood $u=0$.
To avoid a complex metric, we could try matching type I solutions
with type II solutions at $D=0$. This possibility is currently
under investigation \cite{BDG}.

For most of this Section we shall assume $\beta=1$, which ensures
positivity of $D$ for all choices of the potentials $p$ and $q$,
and therefore seems to be the most natural choice
\cite{Isham:1977rj}. For certain potentials, however, there may be
other special values of $\beta$ for which the metric is everywhere
real (as we shall see in Section 3.3).

\subsection{de Sitter with Minkowski}

Let us choose parameters in (\ref{lambafI}-\ref{lambagI}) so that
$\Lambda_g=0$ and $\Lambda_f>0$. Then there is a Type I solution
where $g$ is Minkowski and $f$ is de Sitter. The corresponding
potentials in Eqs. (\ref{spheg}-\ref{sphef}) are given by \be
p=\frac{2\Lambda_f}{9}\ r^2\equiv H^2 r^2, \quad\quad
q=0.\label{potentials}\ee Note that each of the spacetimes,
characterized respectively by the metrics (\ref{spheg}) and
(\ref{sphef}) with the above defined potentials, has a maximal
extension which is geodesically complete (trivial in the case of
Minkowski). However, combining both together will be non-trivial
because the static coordinates $(t,r)$ (where we also include
implicitly the angular part) cover the whole of Minkowski space,
but not the whole of de Sitter.  Hence, the conformal diagram for
the extended de Sitter space accommodates all points for which the
metric $g$ is defined, but the converse is not true. To illustrate
the causal structure, let us represent the light-cones of metric
$g$ in the conformal diagram of $f$. To this end, it is convenient
to use Kruskal-type coordinates, (see e.g. \cite{Gibbons:1977mu})
\be \label{UVdS1} U=-\left(\frac{1-Hr}{1+Hr}\right)^{1/2} e^{-H
\tilde{t}},\quad\quad V=\left(\frac{1-Hr}{1+Hr}\right)^{1/2} e^{H
\tilde{t}}. \ee Note that this involves $\tilde t$ (and not $t$),
the temporal coordinate in which $f$ is diagonal. Eq.
(\ref{UVdS1}) maps the interior of the de Sitter horizon $Hr<1$
into the quadrant $U<0, V>0$ of the plane $(U,V)$. The future
event horizon for an observer at $r=0$ corresponds to $U=0$,
whereas the past event horizon corresponds to $V=0$ (see Fig. 1).
The quadrant $U>0, V>0$ which lies beyond the future event
horizon, is similarly covered by the change of coordinates \be
\label{UVdS2} U=\left(\frac{Hr-1}{Hr+1}\right)^{1/2} e^{-H
\tilde{t}},\quad\quad V=\left(\frac{Hr-1}{Hr+1}\right)^{1/2} e^{H
\tilde{t}}.\ee The remaining quadrants can be obtained by changing
the sign in the right hand side of Eqs. (\ref{UVdS1}-\ref{UVdS2}).
As usual, we may perform the conformal re-scaling $ T=\arctanh\ V+
\arctanh\ U $, and $R=\arctanh\ V - \arctanh\ U$, so that the in
the new coordinates the four quadrants lie in a square of finite
size (see Fig. 1). The vertical boundaries correspond to $r=0$,
while the past and future boundaries of the diagram correspond to
$r=+\infty$ (which is a spacelike boundary). Note further that the
coordinate system $(t,r)$ only covers the $V>0$ corner of the
maximally extended de Sitter spacetime but also that it
accomodates positive and negative values of $U$, so that it goes
beyond the future event horizon. Thus, this coordinate system is
similar, as far as the de Sitter metric is concerned, to the
Eddington-Finkelstein coordinates of a black hole. At this point
one might worry about a possible singularity due to the presence
of the horizon. Indeed, as we discussed above, a coordinate
singularity in one of the two metric cannot always be removed by a
coordinate change that renders both metrics non singular. Here the
situation is different, and in the coordinates ($t,r$), both
metrics are smooth and regular everywhere where $t$ and $r$ take
finite values. So the $U=0$ part of the de Sitter horizon in the
$V>0$ corner does not result in a singularity in the bimetric
theory. Things are however more involved for the $V=0$ part of the
horizon, as we will now see.

\begin{figure}[h]  \centering
\psfrag{r0}[][]{$r=0$}\psfrag{r1}[][]{$r=\infty$}\psfrag{t0}[][]{$t=0$}
\psfrag{U}[][]{$U$}\psfrag{V}[][]{$V$}\psfrag{Vc}[][]{$V=ct.$}\psfrag{r2}[][]{$r=ct.$}
\psfrag{rH}[][]{$r=r_H$}\psfrag{a}[][]{$(a)$}\psfrag{b}[][]{$(b)$}
\includegraphics[width=0.7\textwidth]{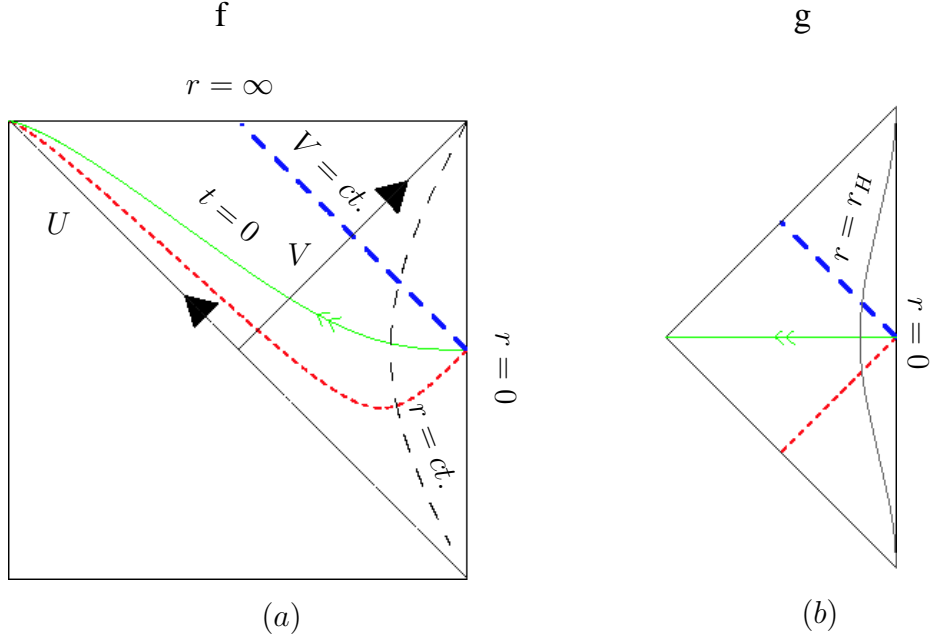}
\caption{\small{Causal diagrams when the $f$ metric is de Sitter
(left diagram) while the $g$ metric is Minkowski (right diagram)
and $\beta=1$. The dashed curly vertical line of the left diagram
represents a sphere of constant radial coordinate $r$. The solid
curly vertical line of the right diagram represents the de Sitter
horizon $r=r_H$ plotted in the Minkowski space-time. We also
plotted three radial geodesics of Minkowksi space-time emanating
from the origin $r=0$ at $t=0$: the thick dashed (blue) curve is a
future-directed radial null ray from the origin (notice it is also
a null geodesic ($V=$ constant) of the  de Sitter space-time), the
thin solid (green) curve with two arrows is a $t=0$ radial
geodesic, the thin dashed (red) curve is a  past-directed null ray
from the origin. The last two curves are radial geodesics of
Minkowski space-time but not of de Sitter space-time. The whole of
the Minkowski space-time is mapped onto the half of the de Sitter
diagram verifying $V>0$. Note that the past directed null
geodesics of Minkowski turn around and start moving towards the
future boundary of de Sitter space. This behaviour, however, does
not lead to closed time-like curves, as discussed in Section
\ref{3.4}}} \label{figure1}
\end{figure}

\begin{figure}[h]  \centering
\psfrag{ip}[][]{$i^+$}\psfrag{im}[][]{$i^-$}\psfrag{imm}[][]{$i^-_{(r<r_H)}$}
\psfrag{imp}[][]{$i^-_{(r>r_H)}$}\psfrag{i0}[][]{$i^0$}
\psfrag{Ip}[][]{$\mathcal{I}^+$}\psfrag{Im}[][]{$\mathcal{I}^-$}
\psfrag{t1}[][]{$t_1$}\psfrag{t2}[][]{$t_2$}
\psfrag{t3}[][]{$t_\epsilon$}
\psfrag{imh}[][]{$i^-_{(r=r_H)}$}
\includegraphics[width=0.7\textwidth]{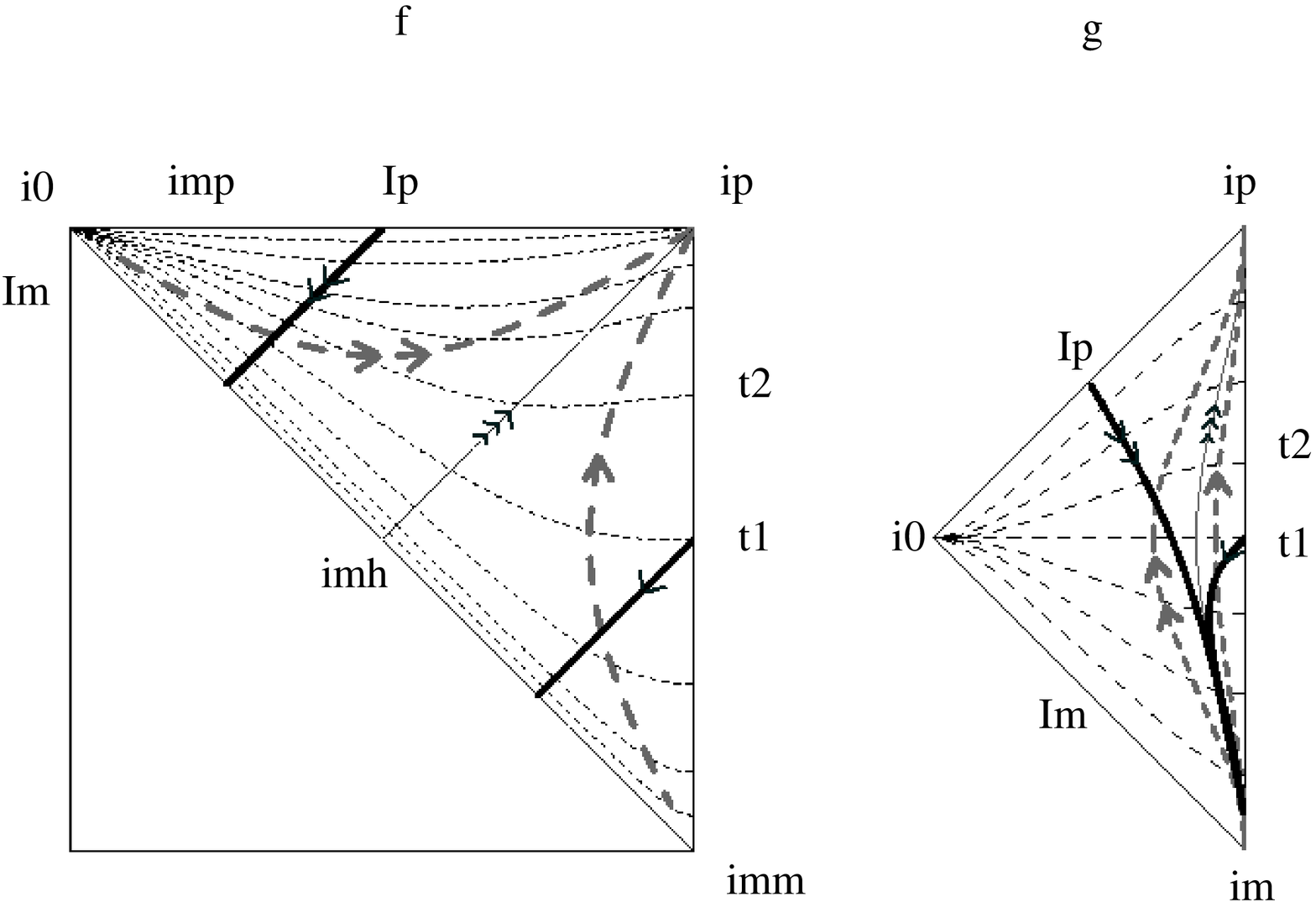}
\caption{\small Causal diagram for de Sitter with Minkowski, for
$\beta=1$. The left diagram is for de Sitter with horizon radius
$r_H$, while the right diagram is for Minkowski. The dashed thin
lines (with no arrows) are $t =$ constant lines. The  dashed thick
line with one (resp. two) arrow is an $r =$ constant curve, with
$r<r_H$ (resp. $r>r_H$). The thin solid line with three arrows
represents the trajectory of an observer sitting at constant
radius $r=r_H$ in Minkowski spacetime. The thick solid lines with
arrows are past directed null geodesics of de Sitter space time
$U= constant$ curves. The mapping of the infinities (null,
spacelike, timelike) of Minkowski spacetimes ($i^{\pm,0}$, ${\cal
I}^{\pm}$) has been indicated on the de Sitter diagram. One of the
stricking feature of those diagrams, is that the past
time-like infinity of Minkowski is split between the upper left
corner (for $r>r_H$), the lower right corner (for $r<r_H$) and the
diagonal ($r=r_H$) of the de Sitter space-time.}
\label{newfigure1}
\end{figure}

To this end, let us consider the light-cones in the Minkowski
metric. Radial null geodesics are simply given by \be \label{MGeo}
t=\epsilon r +k \ee where $\epsilon=\pm 1$ corresponds to future
and past directed null rays respectively. For $\epsilon=0$ we
obtain the space-like $t=k$ slices. In order to represent such
geodesics in the conformal diagram for metric $f$, let us first
express them in terms of $\tilde t$. For the potentials
(\ref{potentials}), Eq. (\ref{ttilde}) reads \be \label{dtilde}
\d\tilde t = \beta^{-1/2} \d t + {H r \over 1-H^2 r^2}
(\beta-1+H^2 r^2)^{1/2}\ \beta^{-1/2} \d r. \ee For $\beta=1$ this
yields \be \label{nullMdS} \tilde{t}= t - r -
\frac{1}{2H}\ln\left|{1-Hr \over 1+ Hr}\right|. \ee The
integration constant has been chosen so that $\tilde t= t$ at
$r=0$. For $\beta\neq 1$, Eq. (\ref{dtilde}) can also be
integrated, but the expressions are a bit more cumbersome and we
shall omit them in what follows. Note that the change of variables
(\ref{nullMdS}) is discontinuous at the de Sitter horizon. This is
just as well, since the coordinates $\tilde t, r$ become singular
at $r \equiv r_H=H^{-1}$,  and we need to consider the
Kruskal-type coordinates anyway. Substituting in (\ref{UVdS1}) or
in (\ref{UVdS2}), we have \be U=\Big(\frac{Hr-1}{Hr+1}\Big)
e^{-H(t - r)}, \quad\quad V= e^{H(t - r)}. \label{change}\ee As
noted above,  these expressions are valid both for $U\leq0$ and
$U\geq0$ (with $V>0$), and so they cover both quadrants
(\ref{UVdS1}) and (\ref{UVdS2}) at once. Now, the radial geodesics
are easily given in the $U,V$ chart (as a curve parametrized by
$r$) by substituting (\ref{MGeo}) into (\ref{change}), \be
U=\Big(\frac{Hr-1}{Hr+1}\Big) e^{-Hk} e^{-H(\epsilon - 1)r},
\quad\quad V=  e^{Hk} e^{H(\epsilon - 1)r}. \label{change2}\ee
Future directed null rays of the Minkowski metric $t=r+k$, are
simply straight lines at 45 degrees,
$$
V= e^{Hk} = const.
$$
On the other hand, past directed null geodesics $\epsilon=-1$, as
well as the spacelike geodesics $\epsilon=0$, have a rather
non-trivial behavior which is illustrated in Fig. \ref{figure1}.
For $Hr \ll 1$, the light-cone emanating from $r=t=0$ (i.e. $k=0$)
has the same shape as in Minkowski space. However, at $Hr \sim 1$
the past directed light-cone opens up and turns around in the
$U,V$ plane. Beyond this turning point, ``past directed" null rays
of Minkowski start progressing towards the future in the de Sitter
diagram! In particular, at large affine parameter, $Hr \to
\infty$, both space-like and past directed null geodesics of
Minkowski meet at the upper left corner of the conformal diagram,
$U\to +\infty, V\to 0$, which belongs to the future boundary of de
Sitter. In fact, the future timelike infinity $i^+$ of Minkowski
is mapped into the upper right corner of the de Sitter
diagram, the future null infinity ${\cal I}^+$ of Minkowski is
mapped into the future null infinity of de Sitter (which is
spacelike), the spacelike infinity $i^0$ and null past infinity
${\cal I}^-$ of Minkowski are both mapped to the upper left corner
of the de Sitter diagram (see figure \ref{newfigure1}). The
situation is more complicated for the past timelike infinity $i^-$
of Minkowski. The latter is split into three pieces: a particle
moving back in time along a $r = constant$ geodesic of Minkowski
space-time would either go to the upper left corner of the
de Sitter diagram if $r > r_H$, to the lower right corner if
$r<r_H$, or to the $U=0, V=0$ central point if $r=r_H$. However, a
given timelike trajectory in Minkowski, stemming from the infinite
past ($t=-\infty, r=r_H$) can emanate in the de Sitter diagram
from any point along the diagonal $V=0$. The latter diagonal is
then representing the whole of the past $r=r_H$ infinity of
Minkowski. This can be better seen, plotting the  null geodesics
of de Sitter into a conformal diagram for Minkowski. Inverting
(\ref{change}), \be t=r + H^{-1} \ln V, \quad\quad
r=\frac{UV+1}{H(1-UV)}, \ee outgoing (or incoming) null curves are
given parametrically in terms of $U$ (or $V$) by taking $V=k$ (or
$U=k$). These are represented in Fig. \ref{newfigure1}. In
particular, one sees that past directed $U=constant$ null lines
can intersect the $V=0$ curve anywhere, while they all asymptote
the $r= r_H$ curve in the Minkowski diagram as $t$ goes to
$-\infty$.

We may then ask whether it is possible to construct a closed
time-like curve by combining signals which propagate in the $f$
metric with those propagating in the $g$ metric. We defer this
discussion to Section 3.4, where we show that this is not possible
for general type I solutions.

\begin{figure}[h]  \centering
\psfrag{a}[][]{$(a)$}\psfrag{b}[][]{$(b)$} \psfrag{D}[][]{$D(r)\in
\mathbb{C}$}
\includegraphics[width=0.7\textwidth ]{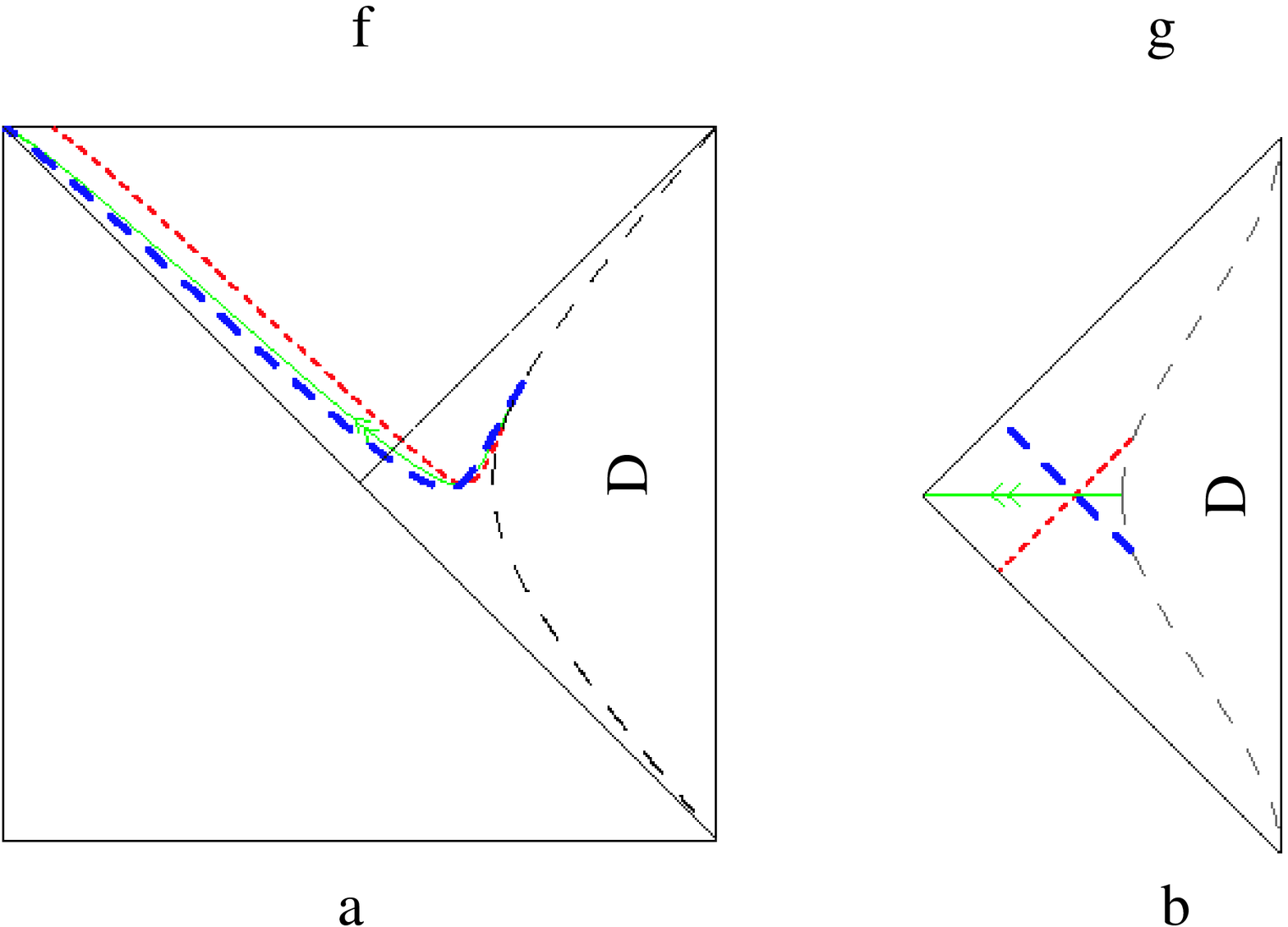}
\caption{\small{ Causal diagrams when the $f$ metric is de Sitter
(left diagram) while the $g$ metric is Minkowski (right diagram)
and $\beta=1/6$. Thick dashed (blue) curve, thin dashed (red)
curve, and thin solid (green) curve with two arrows,  are
respectively null (for the two first) and spacelike (for the last)
radial geodesics of Minkowski space-time. The dashed curly
vertical line in both diagram is an $r=$ constant curve which is
the boundary of the region where one of the metrics becomes
complex.}} \label{figure2}
\end{figure}

A similar analysis can be performed for other values of $\beta$.
For $\beta>1$, $D$ is everywhere real and the causal structure is
quite similar to the one described above. A minor difference is
that the light-cones of Minkowski geodesics are not at 45 degrees
near the origin (as they were in Fig. 1). This can be easily seen
from Eq. (\ref{ttilde}). On the other hand, for $\beta<1$ the
metric becomes complex in the region $H^2r^2 < 1-\beta$ (see Fig.
\ref{figure2}).

Let us now consider the issue of global structure. As was stressed
above, the coordinates $(r,t)$ cover the full Minkowski space
corresponding to the metric $g$, but only half of the conformal
diagram for the extended de Sitter metric, corresponding to $V>0$
(see Fig. 1a). This portion is by itself globally hyperbolic,
since the $t=k$ surfaces are Cauchy surfaces for all geodesics of
both metrics in this region. However, the region $V>0$ is not
geodesically complete, since the the null geodesics $U=const.$ of
de Sitter reach $V=0$ at finite affine parameter. To obtain a
geodesically complete space-time, we can match the solution in the
upper half of the conformal diagram with a solution in the lower
half of the diagram. For this purpose we introduce a {\em second}
Minkowski space, with metric $g'$, which will be covered with
coordinates $r'$ and $t'$. The change of variables (\ref{UVdS1})
and (\ref{UVdS2}) with the substitutions $t\to -t'$, $U\to -U$,
$V\to -V$, maps the full range of the coordinates $r',t'$ into the
lower half of the de Sitter conformal diagram, below the diagonal
$V=0$. The full diagram, represented in Fig. 3, is now
geodesically complete. In doing such an extension,  we mean we are
gluing together one Minkowski spacetime to the other along the
past infinity of the $r=r_H$ sphere of the former to the future
infinity of the $r=r_H$ sphere of the latter. These infinities do
not belong to the Minkowski spacetimes, but to their boundaries,
while they are located in the interior of the de Sitter spacetime.
This provides indeed a perfectly fine geometric maximal extension,
where all geodesics are complete.

\begin{figure}[h]  \centering
\psfrag{a}[][]{$(a)$}\psfrag{b}[][]{$(b)$}
\psfrag{c}[][]{$(c)$}\psfrag{I}[][]{$I$}
\psfrag{II}[][]{$II$}\psfrag{III}[][]{$III$} \psfrag{IV}[][]{$IV$}
\includegraphics[width=0.7\textwidth]{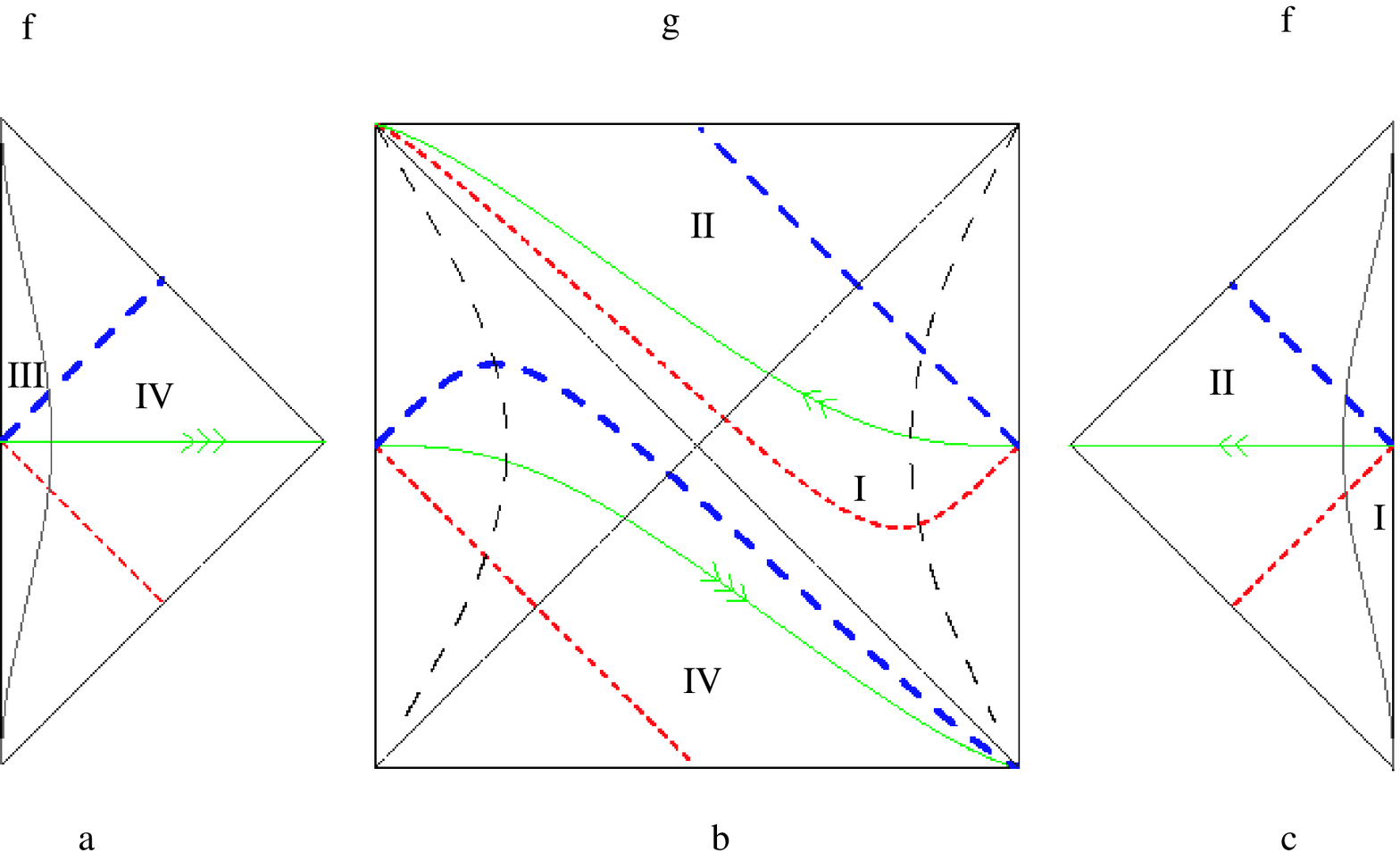}
\caption{\small{Diagram showing the extension proposed in the text
for the de Sitter/Minkowski solution. Notations are the same as in
Fig. \ref{figure1}. By using a second Minkowski space, we can
extend the de Sitter diagram of Fig. \ref{figure1}, represented by
region I and II above,  to the lower half, represented by region
III and IV above. The de Sitter space-time is now geodesically
complete, however the whole space-time it is not globally
hyperbolic, when both metric are considered on the same footing.
If we draw a Cauchy surface for all the de Sitter geodesics [such
as a horizontal line cutting accross the diagram $(b)$], this
surface will intersect some of the Minkowski geodesics twice,
while it will fail to intersect some others.}} \label{figure3}
\end{figure}

\begin{figure}[h]  \centering
\psfrag{a}[][]{$(a)$}\psfrag{b}[][]{$(b)$} \psfrag{I}[][]{$I$}
\psfrag{II}[][]{$II$}\psfrag{III}[][]{$III$} \psfrag{IV}[][]{$IV$}
\includegraphics[width=0.6\textwidth ]{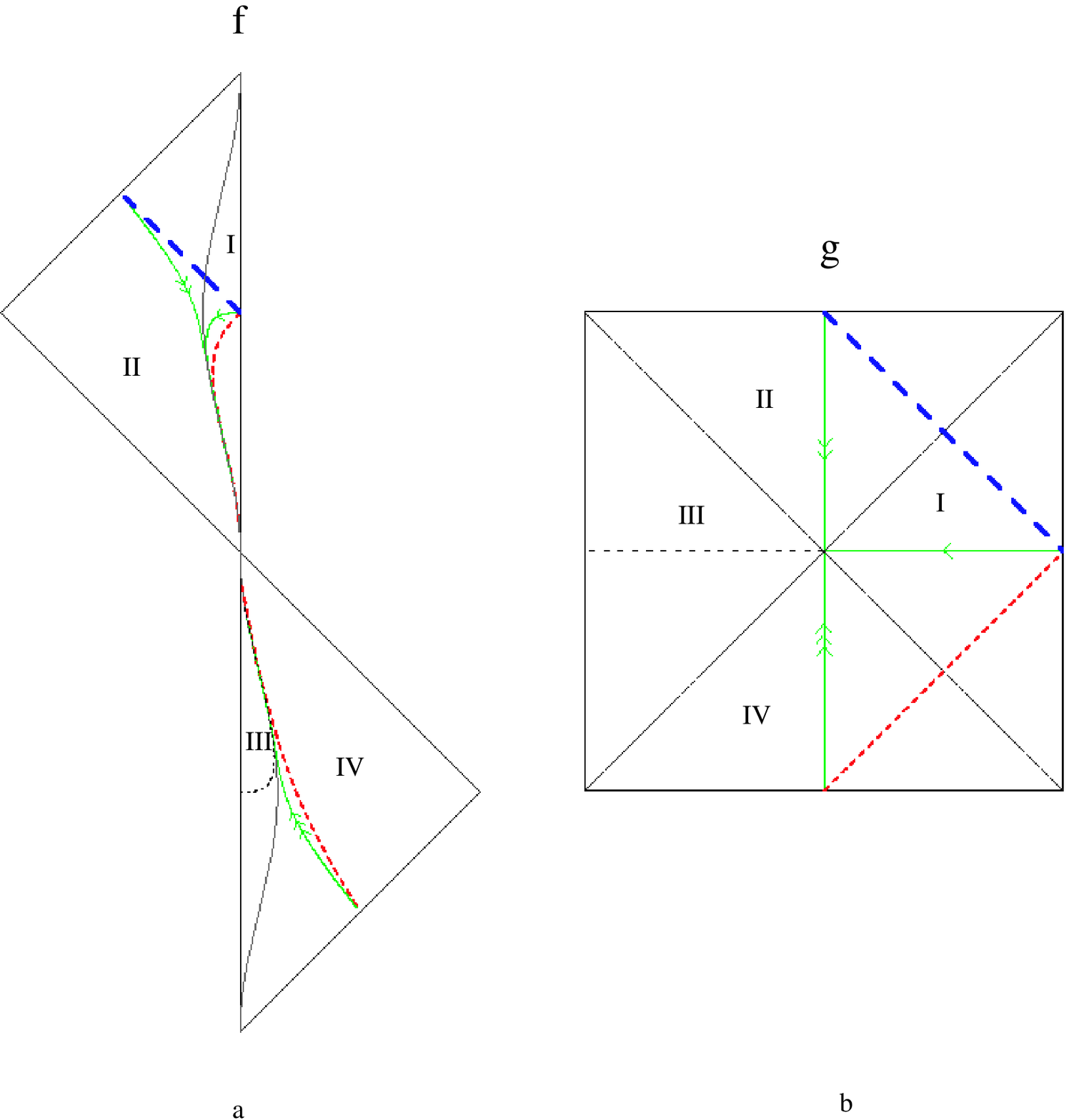}
\caption{\small{Same as Fig. \ref{figure3}, with radial geodesics
of de Sitter plotted instead of those of Minkowski.
 The thick dashed (blue) curve is a future-directed radial null ray from the origin $(r=0, \tilde{t}=0$). The thin solid (green) curve is a $\tilde{t}=0$ radial geodesic of de Sitter.
The thin dashed (red) with one arrow curve a the past-directed
null geodesic from the origin. We also plotted, as thin solid
(green) curves  with two and three arrows, the continuation of the
$\tilde{t}=0$ curve beyond the horizon $r=r_H$.
 When mapped into the Minkowski
diagram, the past directed null geodesics of de Sitter, of region
I, reach the timelike past infinity of the Minkowski space-time at
a finite value of their affine parameter in de Sitter, namely when
they cross the de Sitter horizon $r=r_H$. Nevertheless, we can
"smoothly" continue them in the newly added Minkowski solution
onto which regions III and IV of de Sitter space-time are mapped.
\label{altra}}}
\end{figure}

We should add, however, that a maximal extension is usually
required to satisfy the equations of motion. The bigravity
equations of motion are certainly satisfied everywhere in regions
I, II, III and IV of Fig. 3, but it is unclear in which sense they
are satisfied along the diagonal $V=0$. The problem is precisely
that we are joining two Minkowski spacetimes [$(a)$ and $(c)$ of
Fig. 3] at a locus which lies at their conformal boundary. It is
conceivable that promoting our maximal extension to a solution of
the equations of motion might necessitate additional input, such
as the inclusion of some source at the time-like infinity of
Minkowski.
%Note further, that there is some arbitrariness in
%the extensions which are possible. For example we chose to
%orientate the time in the new Mikowski space time $(t',r')$ such
%that a null geodesic $U =$ constant, of de Sitter space-time,
%crossing the horizon $V=0$ will keep the same (Minkowski) time
%orientation at horizon crossing (otherwise . Along the same line,
%while the extension of the de Sitter part is uniquely defined, it
%is possible to extend the $f$ metric into a Schwarzschild metric
%for example. We will discuss the corresponding Causal diagram in
%the following subsection.
Furthermore, as we shall see below, the extensions are not unique.
This may seem surprising at first sight, but as we shall comment
later on, a similar ambiguity is present in usual General
Relativity when a metric must be continued beyond a Cauchy
horizon.

Note that the  extended diagram, Fig. \ref{figure3}, is not
globally hyperbolic. The $t=k$ surfaces of the region $V>0$ are no
longer Cauchy surfaces for the whole space-time, since they do not
intersect causal geodesics in the lower half of the diagram. A
surface which intersects all causal geodesics should cut through
both regions, $V>0$ as well as $V<0$. One such surface is, for
instance, the horizontal line $U=V$. The problem is that, as can
be seen in Fig. \ref{figure3}, there are geodesics which intersect
this surface twice (such as the past directed null rays from
$r=t=0$).  A formal proof that the maximally extended diagram of
Fig. \ref{figure3} is not globally hyperbolic runs as follows. Let
us restrict attention to radial geodesics. A Cauchy surface must
intersect all causal geodesics once and only once. Let us assume
that such a surface $\Sigma$ exists. In particular, $\Sigma$ must
intersect the null geodesic $V=0$ of de Sitter space. By
continuity, it will also intersect the null geodesics $V=const.$,
in the range $-\delta< V < \delta$, where $\delta$ is an
arbitrarily small positive number. Let us now consider the null
geodesic of Minkowski space, parametrized by $r$ in Eq.
(\ref{change2}), and let us choose the constant $k < H^{-1}
\ln \delta$. It is clear that the incoming radial geodesic (with
$\epsilon=-1$) will start at the upper left corner of the de
Sitter diagram (at $r\to \infty$), and work its way down towards
the right boundary of the diagram (at r=0), while $V$ will always
remain in the interval $0<V<\delta$). Hence, the incoming null
geodesic must intersect $\Sigma$ at least once before it reaches
$r=0$. At $r=0$ it bounces and becomes the outgoing null geodesic
$V=e^{Hk}<\delta$, which will intersect $\Sigma$ once more before
it reaches null future infinity. Hence, there are geodesics of
Minkowski which intersect $\Sigma$ twice, which simply means that
this is not a
good Cauchy surface for all geodesics in the extended diagram.\\

Let us compare the present situation to that in usual GR. As
mentioned above, Cauchy horizons are also present in certain
maximally extended solutions of GR, such as Reissner-Nordstrom or
Anti de Sitter space. Whenever there is such a horizon, the
equations of motion do not suffice to continue the solution past
it, and we need additional input. Usually, analytic continuation
is used, or else some boundary conditions at certain time-like
boundaries of spacetime are introduced. As mentioned above, in the
present context it is not clear whether the equations of motion
are satisfied or not at the Cauchy horizon of the maximally
extended solution, but this is precisely because this horizon
corresponds to a point in the conformal boundary of one of the
metrics. In this sense, the situation is no worse than in GR,
where we have to prescribe data on certain boundaries in order to
determine the maximal extension. Another point to consider is
that, physically, Cauchy horizons tend to be unstable to
perturbations, because of large blueshift effects. The same is
expected to happen in the present context. Note, e.g., from Fig.
\ref{figure3}, that all future directed null geodesics of
Minkowski in regions III and IV tend to pile up near the Cauchy
horizon at $V=0$, suggesting that there will be a large
backreaction near that surface once we include perturbations
\cite{BDG}.

Another interesting fact of the bi-metric solution is that the
concepts of causal past and future are ``broadened", since signals
can be transmitted by matter coupled to both metrics. For
instance, the observers at $r=0$, with $V>0$ can see signals
emitted by all other observers, and hence they have no future
event horizon. Likewise, observers at $r=0$, with $V<0$, can emit
signals which will eventually reach all other observers, and hence
they have no past event horizon. It is tempting to speculate that
cosmological bi-gravity solutions, if they can be made sense of,
could in principle be relevant to the horizon problem.

\subsection{de Sitter with Schwarzschild}

Let us now replace the Minkowski metric by the Schwarzschild one.
In this case, the potentials of the Type I solution are given by
\ba \label{explipq}
 p= H^2 r^2,\quad\quad q=\frac{2 M}{r}. \\ \ea
Both metrics have now horizon singularities whenever $p=1$ and
$q=1$, corresponding respectively to $r=r_H$ and $r=r_S\equiv 2M$.
Those are coordinate singularities from the point of view of each
metric considered separately from the other. However, one might be
concerned by the possibility to remove such singularities from
both metrics at the same time. To study this issue, we first keep
$p$ and $q$ unspecified, and note that the coordinate change
(\ref{ttilde}) reads (with $\beta=1$, which we shall assume in the
following) \footnote{we only discuss here the case
$\epsilon_D=+1$, the other case, which corresponds to a change in
the sign of time, follows similarly} \be \label{trrstar}
d\tilde{t}= \d t - \d r^* + \d\tilde{r}^*, \ee $r^*$ and
$\tilde{r}^*$ defining "tortoise" coordinates associated with
metric $f$ and $g$ respectively by \ba \d r^* &=& \frac{\d r}{1-q}
\label{drs}\\\label{drts} \d\tilde{r}^* &=& \frac{\d r}{1-p}.\ea
Thus, introducing the null coordinates $v = t- r^*, u = t+r^*$ for
the metric $g$, and $\tilde{v}= \tilde{t}-\tilde{r}^*, \tilde{u} =
\tilde{t}+ \tilde{r}^*$, for the metric $f$, one has from the
above expression (\ref{trrstar}) \be \d\tilde{v} = \d v.\ee This
means that $v$ is null for both metrics, but also that
$(v,r,\theta,\phi)$ are Eddington-Finkelstein coordinates for both
metric. In such a coordinates system none of the metric is
singular at the horizons.

Coming back to the explicit expressions for $p$ and $q$
(\ref{explipq}) and substituting those in (\ref{ttilde}) we find
\be \d\tilde{t} = \frac{1}{\sqrt{\beta}}\left\{\d t +
\frac{\sqrt{(H^2r^3-2 M)(H^2 r^3+(\beta-1)r-2\beta
M)}}{(r-2M)(1-H^2r^2)}\d r \right\}, \ee For $\beta=1$, we have
\be \tilde t=t-r^*-\frac{1}{2H}\ln\left|{1-Hr \over
1+Hr}\right|.\ee This matches equation (\ref{trrstar}) where, the
Schwarzschild ``tortoise" coordinate reads
\begin{equation}
r^*=r+2M \ln|1-r/2M|.
\label{tortoise}
\end{equation}
The analog of Eq. (\ref{change}) is now \be
U=\Big(\frac{Hr-1}{Hr+1}\Big) e^{-H(t - r^*)}, \quad\quad V=
e^{H(t - r^*)}, \label{changesch}\ee which, again, is valid both
for $U>0$ and $U<0$ (with $V>0$), covering both quadrants
(\ref{UVdS1}) and (\ref{UVdS2}) of de Sitter,  that is to say the
region covered by the Eddington-Finkelstein coordinates $(v,r,
\theta, \phi)$. The null and spacelike radial geodesics of
Schwarzschild can be written as \be t= \epsilon r^*+k,\label{geo}
\ee  this being obviously valid in the whole region covered by
coordinates $(v,r, \theta, \phi)$.
%Note that the Schwarzschild
%metric is smooth near the horizon when written in terms of the
%null coordinates $t\pm r^*$. Therefore, even though the relation
%(\ref{tortoise}) is singular at $r=2M$, the value of the constant
%$k$ in (\ref{geo}) is continuous accross the horizon.
In the $U,V$ chart these geodesics are given by \be
U=\Big(\frac{Hr-1}{Hr-1}\Big) e^{-Hk} e^{-H(\epsilon - 1)r^*},
\quad\quad V=  e^{Hk} e^{H(\epsilon - 1)r^*}. \label{change22}\ee
Again, we find that the null geodesics $t=r^*$ correspond to
$V=const.$, (or $v = const$) so $V$ is a null coordinate both in
Schwarzschild and in de Sitter. The other radial geodesics, with
$\epsilon=-1,0$ have a more complicated form, which is
qualitatively represented in Fig. \ref{SCHDSmatched}. Note that for this
figure, we have assumed that the Schwarzschild radius $r_S$ is
smaller that the de Sitter horizon radius $r_H$.

\begin{figure}[h]  \centering
\psfrag{ip}[][]{$i^+$}\psfrag{im}[][]{$i^-$}\psfrag{imm}[][]{$i^-_{(r<r_H)}$}
\psfrag{imp}[][]{$i^-_{(r>r_H)}$}\psfrag{imh}[][]{$i^-_{(r=r_H)}$}
\psfrag{iz}[][]{$i^0$}\psfrag{r0}[][]{$r=0$}
\psfrag{Ip}[][]{$\mathcal{I}^+$}\psfrag{Im}[][]{$\mathcal{I}^-$}
\includegraphics[width=0.7\textwidth ]{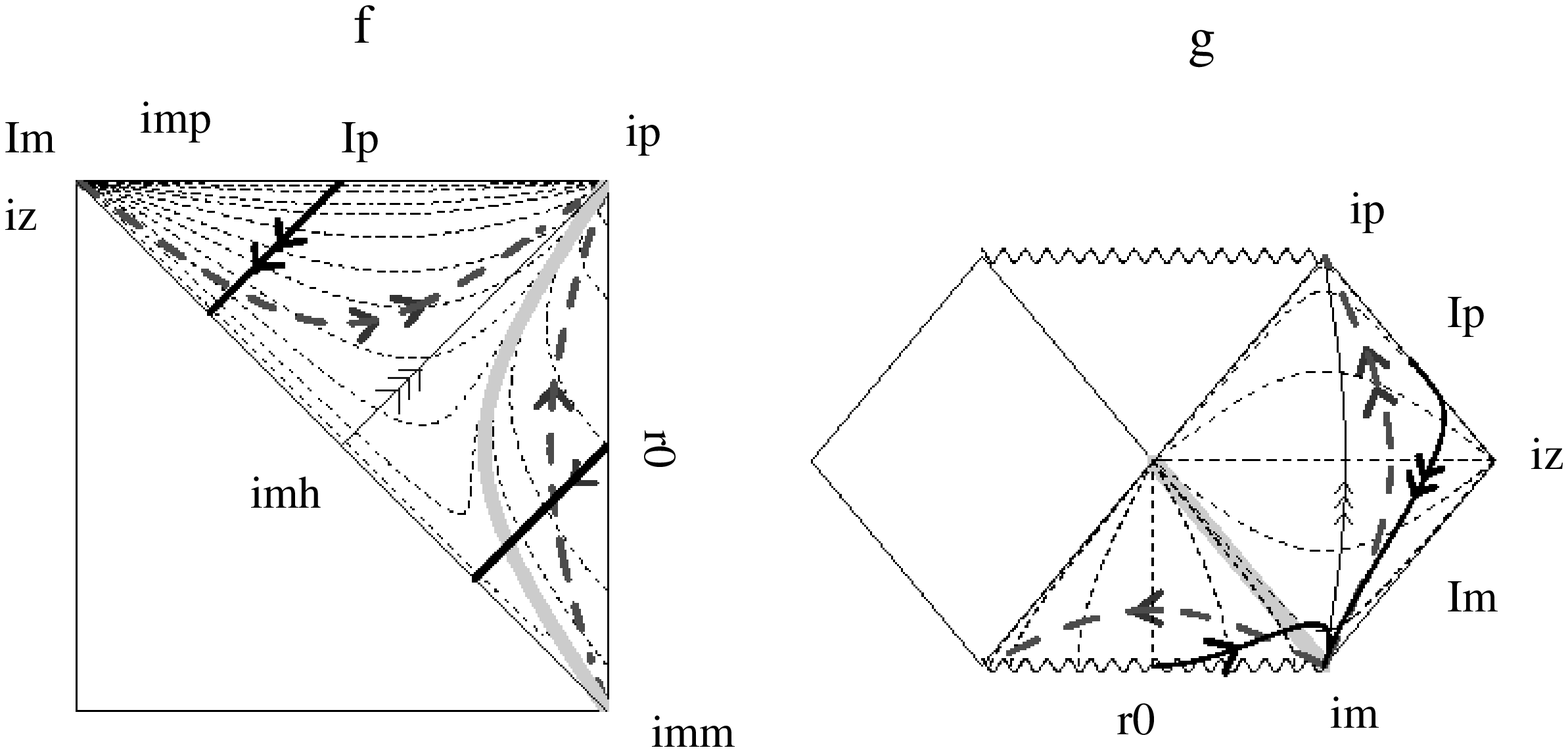}
\caption{\small{Causal diagrams when the $f$ metric is de Sitter (right)
and the $g$ metric is Schwarzschild (left). The notations are the same
as in figure \ref{newfigure1}. The main difference with the case depicted
in this last figure is the presence of the Schwarzschild horizon. The part of
the Schwarzschild horizon shown as a thick gray line on the right diagram above
is mapped to the thick gray line of the left diagram.
The part of the Schwarzschild horizon which is the diagonal of the right
diagram orthogonal to the thick gray line is mapped to the upper right corner of the
de Sitter diagram in analogy to what was found to happen for the de Sitter horizon when the other metric is Minkowski. This shows the possibility to extend the Schwarzschild space-time through another de Sitter spacetime joined to the other by the future infinity of a $r=r_S$ sphere ($r_S$ being the Scharzschild horizon)\label{SCHDSmat}}}
\end{figure}

\begin{figure}[h]  \centering
\psfrag{a}[][]{$(a)$}\psfrag{b}[][]{$(b)$} \psfrag{I}[][]{$I$}
\psfrag{II}[][]{$II$}\psfrag{III}[][]{$III$} \psfrag{IV}[][]{$IV$}
\includegraphics[width=0.7\textwidth ]{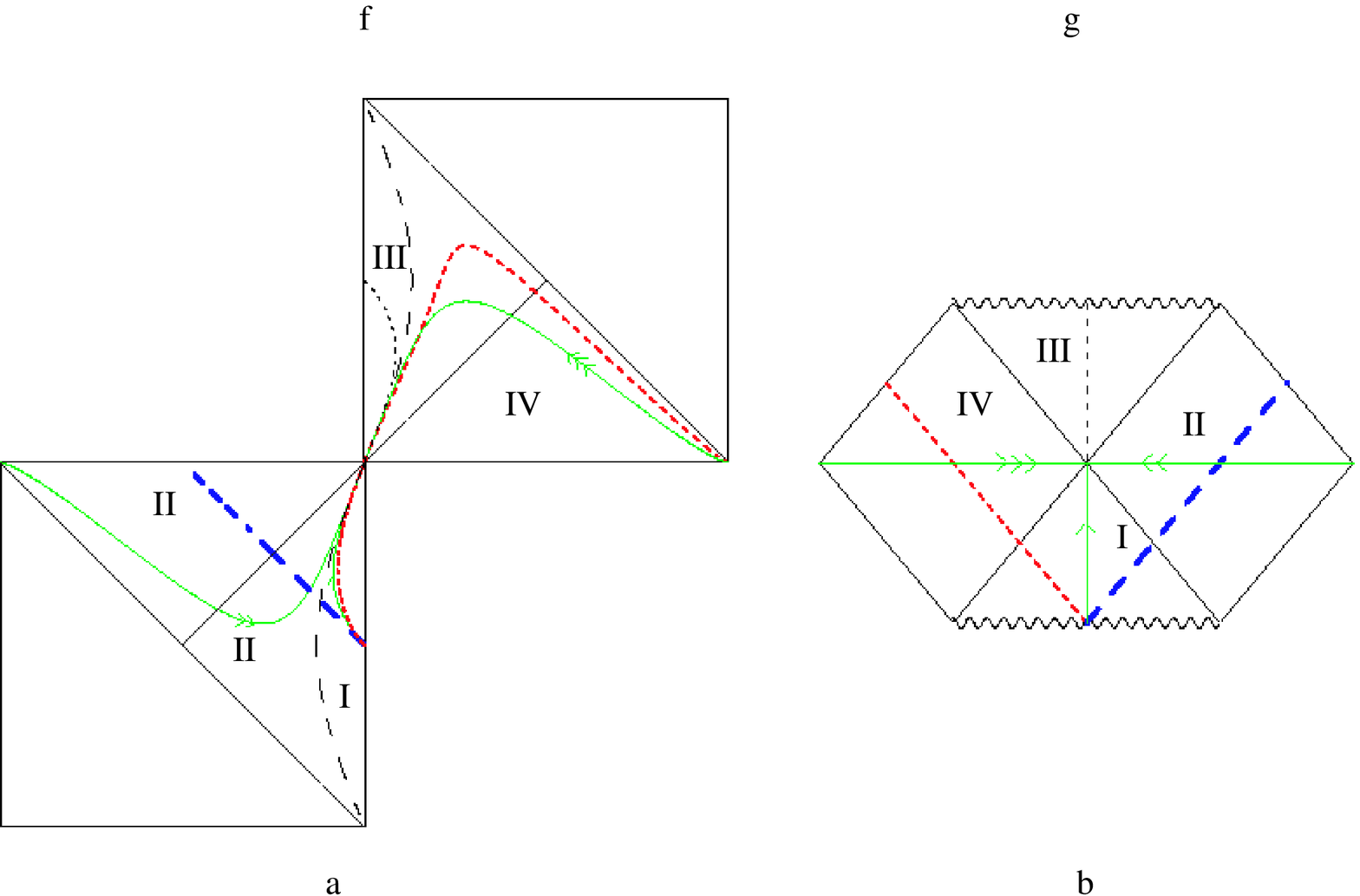}
\caption{\small{Causal diagrams when the $f$ metric is de Sitter (right) and the $g$ metric is Schwarzschild (left) showing the extension proposed in the text for the Schwarzschild space-time. Various radial geodesics of Schwarzschild are mapped onto the de Sitter diagram The dashed vertical curly line in
the de Sitter diagrams indicates the Schwarzschild horizon.
 Note that we can
"send a signal" from region I of the lower de Sitter space to
region IV of the upper de Sitter space by using the left-moving
null geodesic of Schwarzschild (thin dashed (red) line). \label{SCHDSmatched} }}
\end{figure}

As we discussed previously, and is manifest from Fig.
\ref{SCHDSmat}, half of the de Sitter
diagram (above the diagonal) is mapped onto half of the
Schwarzschild diagram (below the diagonal), corresponding to the
region mapped by the Eddington-Finkelstein coordinates
$(v,r,\theta, \phi)$. Both half-diagrams are geodesically
incomplete, since some geodesics reach the horizons (which
dissects the diagrams in two) at finite affine parameter. These
geodesics can of course be extended by adding new regions of
spacetime. If one adds de Sitter and Schwarzschild regions, one
obtains a "stair-case" diagram with an infinite chain of de Sitter
and Schwarzschild space-times, two adjacent de Sitter (resp.
Schwarzschild) space-times being linked together by a common
Schwarzschild (resp. de Sitter) space-time. Needless to say, there
is also a tension in this case between geodesic completeness and
global hyperbolicity, as we found in the Minkowski-de Sitter case.

This seems to apply to more general situations where one of the
metrics has an horizon which is not shared by the other one
\cite{BDG}. As noted previously, the new metric (new "step")
which can be added to the stair does not necessarily correspond to
the same solution as the one of the last step of the stair, since one of the two metrics does not determine
uniquely the form of the other. Thus, in general we can construct "stair-case" diagrams with steps having different forms. Note
further, that in the case considered here, the stairs can always be finished by adding a Minkowski spacetime, linked to a Schwarzschild space-time along a sphere of radius $r_H$ at time-like infinity.

\begin{figure}[h]  \centering
\psfrag{a}[][]{$(a)$}\psfrag{b}[][]{$(b)$} \psfrag{I}[][]{$I$}
\psfrag{II}[][]{$II$}\psfrag{III}[][]{$III$} \psfrag{IV}[][]{$IV$}
\includegraphics[width=0.7\textwidth ]{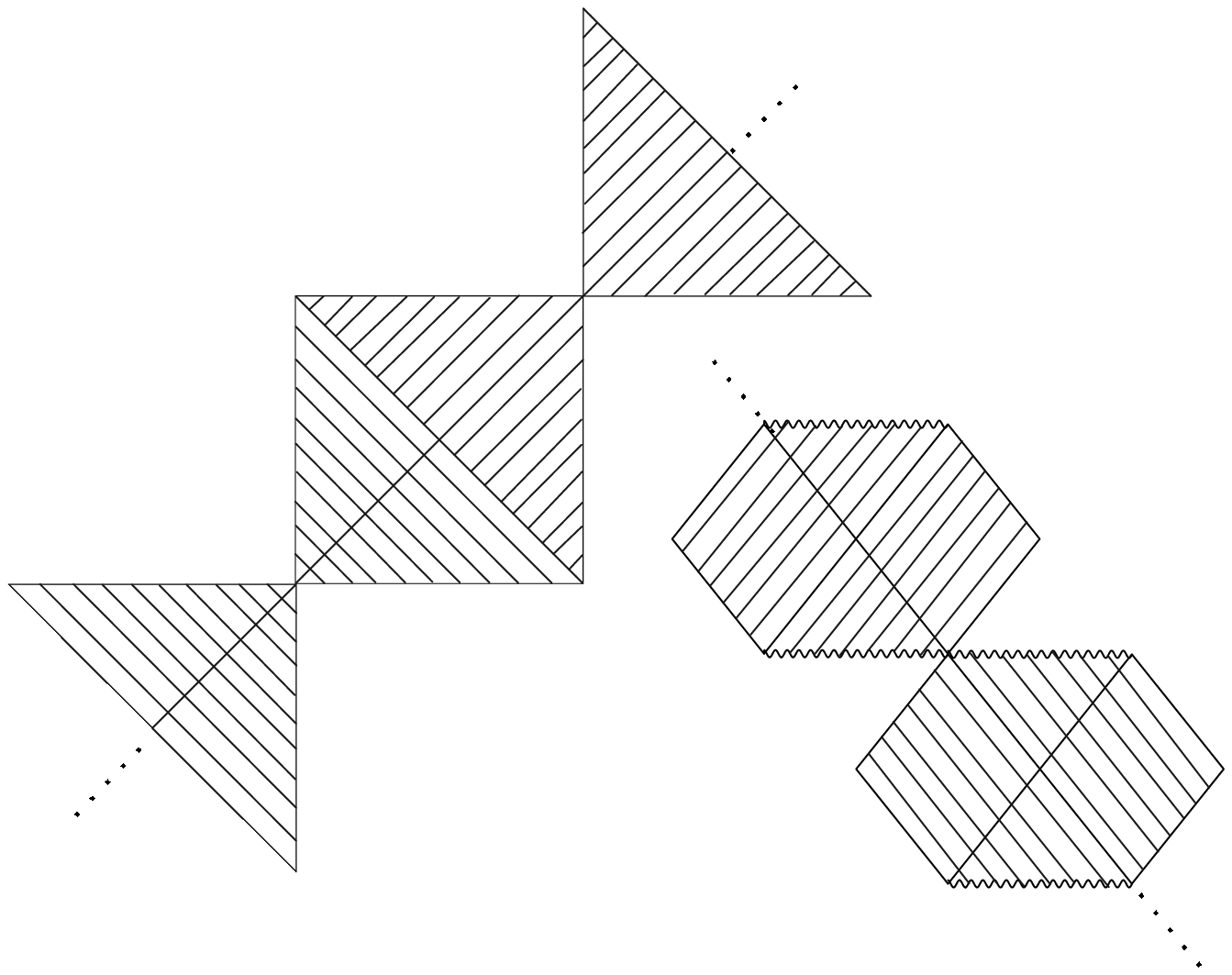}
\caption{\small{This shows a possible maximal extension of the bi-metric space-times, following the procedure given in the text, when one of the metric is de Sitter while the other is Schwarzschild. We are led to the "stair-case" diagram, an
infinite chain of de Sitter spaces linked to each other through a
common Schwarzschild diagram.\label{Stairs}}}
\end{figure}

\subsection{de Sitter with de Sitter}

\begin{figure}[h]  \centering
\psfrag{a}[][]{$(a)$}\psfrag{b}[][]{$(b)$} \psfrag{I}[][]{$I$}
\psfrag{II}[][]{$II$}\psfrag{III}[][]{$III$} \psfrag{IV}[][]{$IV$}
\includegraphics[width=0.7\textwidth ]{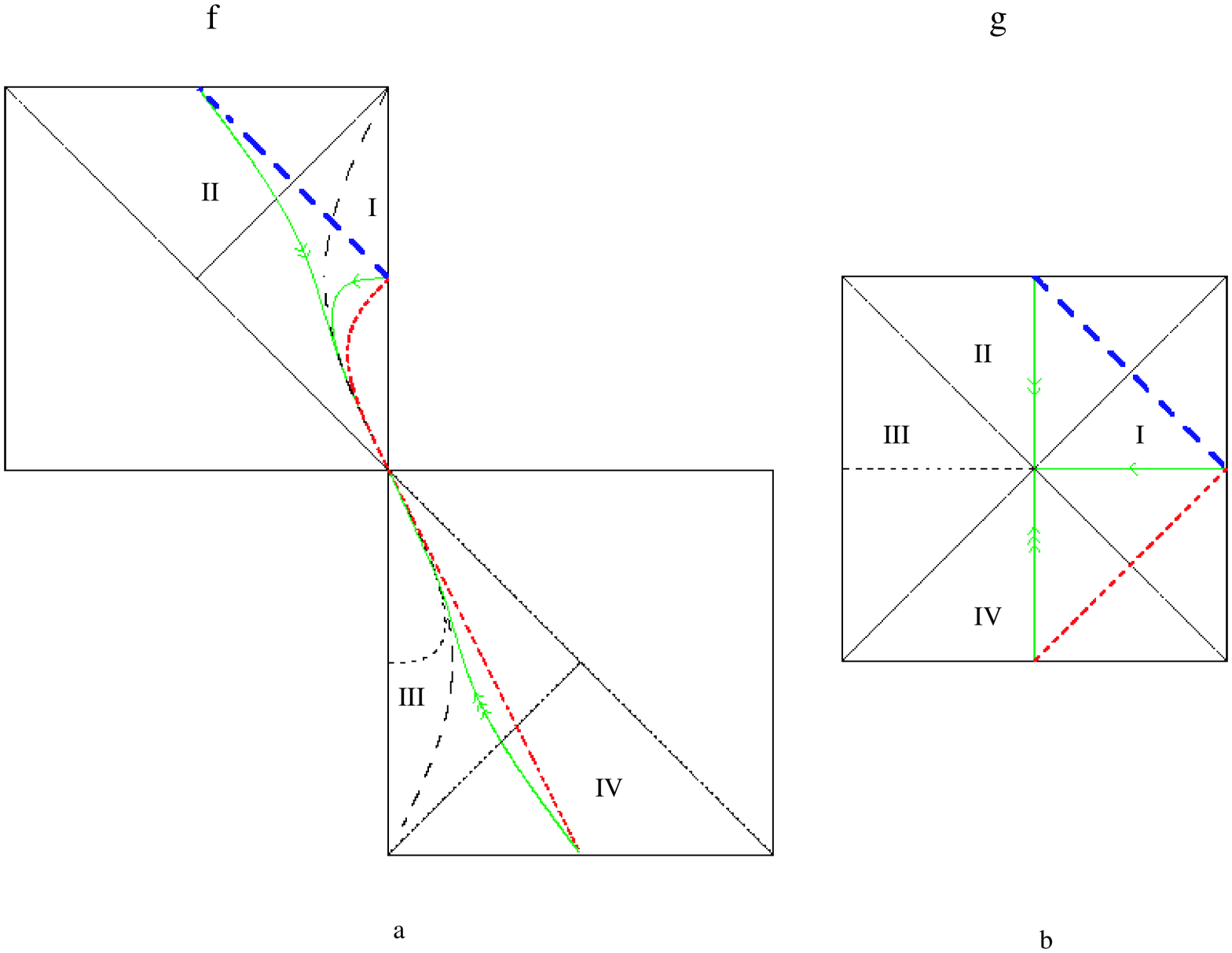}
\caption{\small{Causal diagram when both metric are de Sitter and $\beta=1$.
Notations are the same as in figure \ref{altra}.}\label{dSdS}}
\end{figure}

When both metrics are de Sitter, the potentials are given by \be
p= H_1^2 r^2  \quad\quad q=H^2_2 r^2. \ee For $\beta=1$, the
analysis proceeds along the same lines as in the previous
Subsection, with the only difference that the (de Sitter) tortoise
coordinate is now given by
\begin{equation}
r^* = -{1\over 2H_2} \ln\left|{1-H_2 r \over 1+ H_2 r}\right|.
\label{torto}
\end{equation}
The corresponding causal diagram is represented in Fig. \ref{dSdS}

\begin{figure}[h]  \centering
\psfrag{a}[][]{$(a)$}\psfrag{b}[][]{$(b)$} \psfrag{I}[][]{$I$}
\psfrag{II}[][]{$II$}\psfrag{III}[][]{$III$} \psfrag{IV}[][]{$IV$}
\includegraphics[width=0.7\textwidth ]{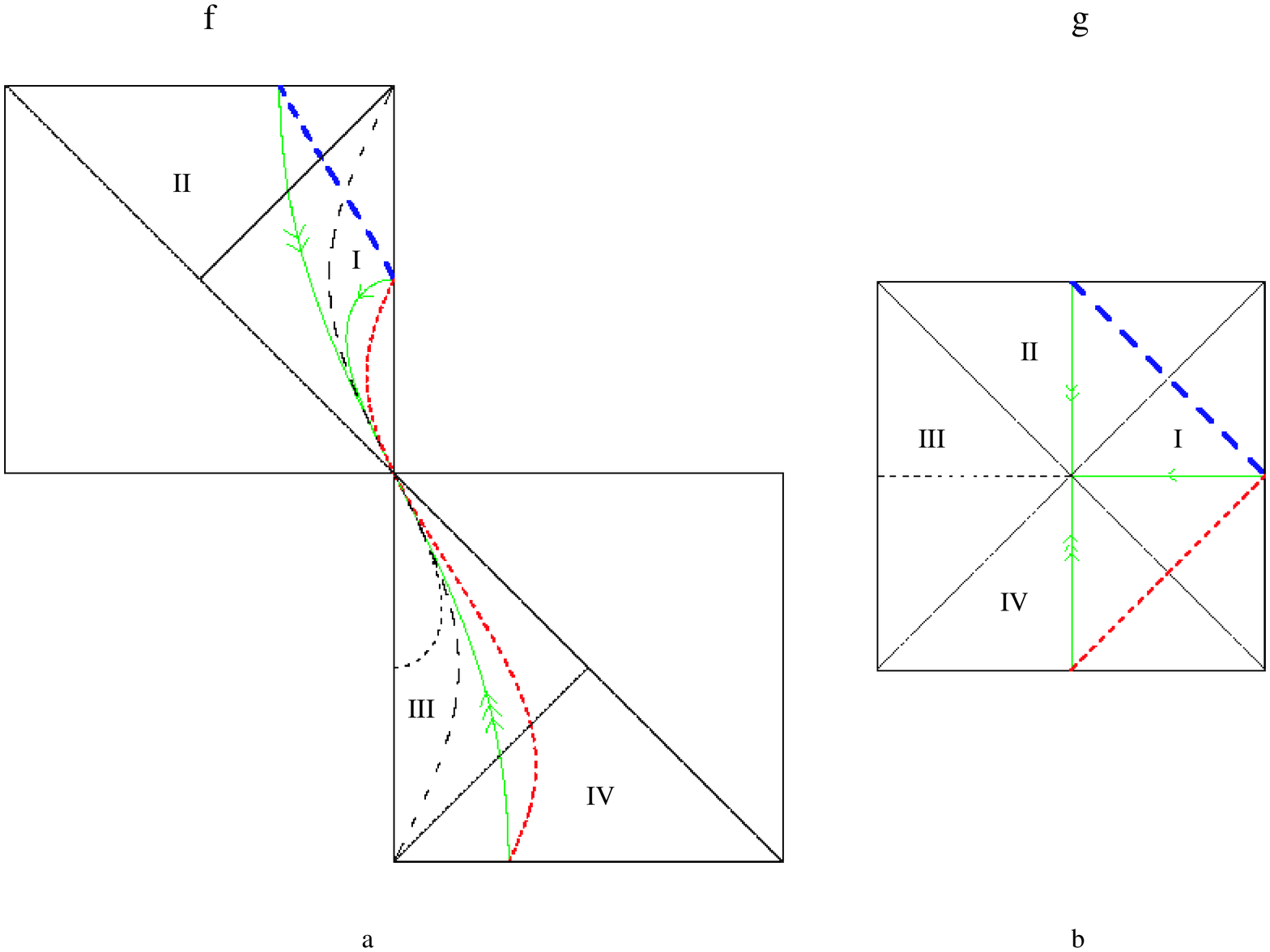}
\caption{\small{Causal diagram when both metric are de Sitter and $\beta=1/4$. Notations are the same as in figure \ref{altra}.}\label{dSdSbeta}}
\end{figure}

Aside from the choice $\beta=1$, the de Sitter de Sitter solution
allows for another way of having $D^2>0$ for the entire range of
$r$.  Indeed, it is enough to have $H_1^2 \geq \beta H_2^2$ and
$\beta \geq 1$ or $H_1^2 \leq \beta H_2^2$ and $\beta \leq 1$.
Choosing for example $\beta$ given by
\begin{equation}
\beta= {H_1^2\over H_2^2},
\end{equation}
we have \be H_1\tilde t = H_2 t - \frac{1}{2}\ln \left|1-H_1^2r^2
\over 1-H_2^2r^2\right|, \ee or $H_1(\tilde t -\tilde
r^*)+\ln(1+H_1 r) = H_2 (t-r^*)+\ln(1+H_2 r)$. Thus, the Kruskal
coordinates (\ref{UVdS1}-\ref{UVdS2}) for the metric $p$ can be
expressed in terms of coordinates $t$ and $r$ as \be
U=\left(\frac{H_1r-1}{H_2r+1}\right) e^{-H_2 (t-r^*)},\quad\quad
V= \left({H_2 r + 1 \over H_1 r +1}\right) e^{+H_2 (t-r^*)},\ee
where $r^*$ is given by (\ref{torto}). The corresponding diagram
is given in Fig. \ref{dSdSbeta}.

\subsection{Closed time-like curves?}
\label{3.4} An interesting question regarding the bigravity
solutions is whether we can construct closed time-like curves by
patching together geodesics corresponding to both metrics.

For $\beta=1$ it is easy to show that this cannot be done, by
using the "tortoise" coordinates $r^*$ and $\tilde{r}^*$ that we
defined in equations (\ref{drs}) and (\ref{drts}), as well as the
null (for both metric) coordinate $v$ (in all this subsection, we
keep the functions $p$ and $q$ unspecified). The radial null and
time-like geodesics of both metrics are given by
$$
t=\epsilon r^* + k,\quad\quad \tilde t= \tilde\epsilon \tilde
r^*+\tilde k,
$$
(Here $\epsilon=\pm 1, 0$ for outgoing and incoming null rays, or
for spacelike geodesics, respectively, and similarly for
$\tilde\epsilon$).
%, which shows that $w$ is a null coordinate in both metrics.
Thus,  any future directed causal curve with respect to $f$ or $g$
has the property that $\d v \geq 0$, and $\d v$ vanishes only
along the outgoing null radial geodesic. Once $v$ increases, even
if it is by just a little bit, it is impossible to go back to the
original value by following a future directed time-like curve,
which means that such curve cannot be closed.

Here, we disregard the possibility of making global
identifications in the coordinate $v$, which might allow for the
construction of a closed loop. Of course, even in flat space with
a single metric, closed time-like curves could be constructed by
global identifications, and in what follows we shall ignore this
somewhat artificial setup. We shall only be concerned with the
possibility of locally constructing closed time-like curves within
a given coordinate patch of space-time, without identifications.

To analyse the general case $\beta\neq 1$ it is convenient to
separately consider the following regions of space-time:

{\em a:} For $(1-p)<0$, and $(1-q)<0$ the condition $\d r=0$ defines
a space-like surface for both metrics $f$ and $g$. This means that
$r$ can only change monotonically along time-like curves of both
metrics, making it impossible to close them in this region.

{\em b:} For $(1-p)<0$ and
$(1-q)>0$, the condition $\d t=0$ defines a space-like surface for
the metric $g$. Also, from (\ref{ttilde}) with $\d t=0$, we have
\begin{equation}
\left|{d\tilde t \over d\tilde r^*}\right|^2 =1+{1\over \beta}
\left({1-p \over 1-q}\right)^2-{\beta+1\over\beta}\left({1-p \over
1-q}\right)
> 1.
\label{puri}
\end{equation}
Since $\tilde t$ is space-like in metric $f$ this means that the
surface $\d t=0$ [which is also defined by Eq. (\ref{puri})] is
space-like in metric $f$ too. Hence, $t$ changes monotonically
along time-like curves of both $f$ and $g$, and as a consequence
such curves cannot be closed.

{\em c:} If $(1-p)>0$ and $(1-q)<0$, then the surface $\d\tilde
t=0$ is space-like for $f$. From (\ref{ttilde}) with $\d\tilde
t=0$, we have
\begin{equation}
\left|{\d t \over \d r^*}\right|^2 =1+\beta \left({1-q \over
1-p}\right)^2-(\beta+1)\left({1-q \over 1-p}\right)
> 1.
\label{pili}
\end{equation}
Since $t$ is space-like in metric $g$, Eq. (\ref{pili}) means that
the surface $\d\tilde t=0$ is space-like in metric $g$ too, and
$\tilde t$ must be monotonic on time-like curves, which therefore
cannot close.

{\em d:} Finally, if $(1-p)>0$ and $(1-q)>0$, then  we must
distinguish two cases. For $p \geq q$, it is easy to see that
$A>0$ in Eq. (\ref{sphef}), and therefore $\d t=0$ is space-like for
both metrics $f$ and $g$. Hence, $t$ is monotonic for time-like
curves of both metrics. On the other hand, for $p\leq q$, Eq.
(\ref{pili}) for $\d\tilde t=0$ leads to
\begin{equation}
\left|{\d t \over \d r^*}\right|^2 < 1.
\end{equation}
Since now $t$ is time-like in metric $g$, this means that $\d\tilde
t=0$ is a space-like surface for this metric. Of course $\d\tilde
t=0$ is also space-like for $f$, and so $\tilde t$ is monotonic
along causal curves for both metrics.

This completes the proof for the individual regions listed above.
It is remarkable that in spite of the strong differences in the
light-cone structure of both metrics, it is not possible to draw
closed time-like curves in any of the regions. The reason is that
the future light-cone for one of the metrics never contains a part
of the past light-cone for the other metric. Thus, we can always
find a coordinate which labels hypersurfaces which are space-like
for both metrics. This coordinate must grow monotonically along
time-like curves.

By continuity, at the boundaries in between the regions, the
future light-cone of one of the metrics can at most touch the past
light-cone of the other metric, sharing perhaps a common null
direction for both metrics. Even if this were the case, a future
directed time-like geodesic with respect to one of the metrics can
never get to the inside of the past light cone with respect to the
other metric, and closed time-like curves cannot be constructed
even if we cross the boundaries between the individual regions.
\footnote{In the examples we have examined, the situation where
the future light-cone of one of the metrics marginally touches the
past light-cone of the other metric at the boundary between
regions does not arise. If it did, then there might closed
future-directed {\em null} curves at such boundary. Note, however,
that since the boundary is at $r=const.$, this situation can only
happen when both metrics have a common event horizon at the same
value of $r$. The possibility of having closed null curves on
these boundaries may require a case by case analysis, and is left
for further research.}

\section{Conclusions}

In this paper we have analyzed the global structure of a wide
class of spherically symmetric bigravity solutions. We have
reviewed their classification, pointing out that vacuum energy
terms of the matter lagrangian can be included in type I
solutions, and we have discussed a new subclass of type II
solutions.

In order to probe the global structure of the solutions, the basic
strategy has been to map the radial null geodesics of one of the
metrics (say $g$) in the conformal diagram of the companion
geometry (corresponding to metric $f$). The relevance of such
geodesics is two-fold. First of all, they are followed by test
particles of the matter sector coupled to each metric. Second, in
the limit of high frequency, the interaction term between $f$ and
$g$ is negligible compared with the kinetic term, and short
wavelength gravitons will travel along the light cones of the
corresponding metrics.

The results illustrate several intriguing features, originating in
the different light-cone structure of metrics $f$ and $g$. For
instance, a {\em past directed} null geodesic of metric $g$ may
well end up at the {\em future} boundary of $f$, as happens in the
case when $f$ is de Sitter and $g$ is Minkowski (Fig. 1). In spite
of this seemingly bizarre behaviour, we have also shown that for
all known solutions, it is not possible to construct closed
time-like curves by piecing together geodesics of both metrics.
The basic reason is that the future light cone with respect to one
of the metrics never invades the interior of the past light cone
with respect to the other metric.

Another peculiarity is that, even if metric $f$ is geodesically
complete by itself, some of its geodesics reach the conformal
boundary of $g$ at a finite value of their affine parameter. In
order to extend them, further "copies" of the full conformal
diagram of $g$ can be incorporated. In some cases, this
leads to an infinite chain of diagrams, such as the "stair-case"
of Fig. \ref{Stairs}, where adjacent de Sitter diagrams
(corresponding to metric $f$) are linked together through a common
Schwarzschild diagram (corresponding to metric $g$). In this
maximally extended bigravity solution, one could send a message
from one de Sitter space to the next by travelling along a
geodesic of metric $g$.

The inclusion of several copies of the conformal diagram of a
given metric leads, in general, to lack of global hyperbolicity,
even when the "building block" diagrams are globally hyperbolic
with respect to their own geodesics. Note, however, that global
hyperbolicity breaks down in many exact solutions of ordinary
General Relativity (GR), such as Reissner-Nordstrom or in Anti-de
Sitter space. This is not necessarily a problem for practical
applications, because usually only a part of the diagram will be
relevant.

The concept of event horizon is broadened in the context of
bi-metric solutions. In principle, through gravitational
interactions, any observer will be affected at some level by
signals propagating in either metric. In particular, a geodesic
observer sitting at $r=0$ in de Sitter space will have no future
event horizon, provided that he or she is sensitive to the signals
which are sent along the geodesics of Minkowski (see Fig. 1).

In summary, the global structure of bimetric solutions presents
some unusual features which, nevertheless, do not seem more
pathological than those encountered in maximally extended
solutions of GR. It is interesting that all the bi-metric
solutions considered are such that both metrics $f$ and $g$ are
solutions of Einstein's equations with a cosmological term. Their
possible phenomenological relevance seems therefore worthy of
further consideration, and will be discussed elsewhere \cite{BDG} (see also ref.
\cite{Damour:2002wu}).

\section{Acknowledgements}
C.D. thanks the Departament de F\'{\i}sica Fonamental of the
Universitat de Barcelona for its hospitality while this work was
initiated. D.B. and J.G. thank the Institute d'Astrophysique de
Paris for its hospitality. We thank S.~Carlip, B.~Carter,
R.~Emparan, E.~Flanagan, V.~Frolov and T.~Jacobson for
discussions. The work of D.B. has been supported by MEC (Spain)
through a FPU grant. The work of J.G. is supported by grants FPA
2004-04582-C02-02 and DURSI 2001-SGR-0061.

\section*{Appendix A: Type II solutions}

Here we show that the most general Type II solution where one of the
metrics satisfies (\ref{Einstein}), is such that $f_{\mu\nu}= \gamma
g_{\mu\nu}$, where $\gamma$ is a constant whose value is given by
the equations of motion.

From (\ref{Einstein}), we have $K T^g_{tt} + J T^g_{rr}=0$, and
plugging expressions (\ref{formg}) and (\ref{formf}) into Eqs.
(\ref{Tga}), we have \ba \label{Tg1lin} K T^g_{tt} + J T^g_{rr}
&=& \frac{\zeta B}{2 r^4} \left(\frac{\Delta B^2
}{JKr^4}\right)^{v-1} (AJ-CK)(3B-2r^2)=0. \ea Since we are now
assuming that $B \neq (2/3) r^2$, it follows that \be
\label{casetypeII} AJ-CK = 0. \ee Hence, from (\ref{formg}) and
(\ref{formf}) plugged into (\ref{Tf}),
 \ba \label{Tf1lin} A T^f_{tt} + C T^f_{rr} &=& -
\frac{\zeta }{2 B} \left(\frac{JK r^4}{\Delta
B^2}\right)^{u}(AJ-CK)(3B-2r^2) =0, \ea and from the equations of
motion \be A R^f_{tt} + C R^f_{rr} = 0. \ee From this we obtain
(see e.g. \cite{Isham:1977rj} for the explicit expressions of the
Ricci tensor components), \be -B^{\prime \prime} + \frac{B^{\prime
2}}{2B} + \frac{\Delta^\prime B^\prime}{2 \Delta}=0. \ee A first
integral is given by \be \label{eqdiffB} \frac{B^{\prime 2}}{B}=4
a^2 \Delta \ee where $a$ is the constant of integration.

Let us now consider the linear
combination \ba r^2 T^g_{tt} + J T^g_{\theta \theta} &=&
-\frac{\zeta}{2K r^2} \left(\frac{AB^2
C}{JKr^4}\right)^{v-1}(BJ-Cr^2)(BK-3AB+Ar^2), \ea which again must
vanish if $g$ is a is a solution of (\ref{Einstein}). Thus one
either has \be \label{case2} BK
+ A r^2 = 3 AB, \ee or \be \label{case1} BJ = C r^2.\ee In both cases
\ba B T^f_{tt} + C T^f_{\theta
\theta} &=& \frac{\zeta}{2 AB} \left(\frac{JKr^4}{AB^2C}\right)^u
(BJ-Cr^2)(BK-3AB+Ar^2)=0.\label{cath} \ea
Note that (\ref{Tf1lin}) and (\ref{cath}) imply that $T^f_{\mu\nu}=H(r)f_{\mu\nu}$. The equations of motion require that $T^f$ must be covariantly conserved, which implies that $H$ is a constant. Therefore, $f$ is a solution of Einstein's equations with a cosmological constant.

Consider first the case when (\ref{case2}) is satisfied. From
this equation and (\ref{casetypeII}), we can eliminate $A$ and $C$
as functions of $B$ and $J=K^{-1}$. We get from (\ref{eqdiffB})
\be \label{second} \frac{B^{\prime 2}}{B^3} = \frac{4 a^2}{(3B -
r^2)^2}. \ee
With the change of variable
\be
B(r)=r^2F^2(r),
\ee
the differential equation (\ref{eqdiffB}) is written as
\be
 r F^{\prime}= \frac{a F^2}{(3F^2 -1)}-F,
\ee
which can be easily integrated to give
\be
c r = {1\over F}\Big(\frac{\sqrt{12+a^2}-a+6F}{\sqrt{12+a^2}+a-6F}\Big)^{\frac{a}{\sqrt{12+a^2}}},\label{tess}
\ee
where $c$ is an integration constant. Notice that
\be
F(r)=(\sqrt{12+a^2}+a)/6, \label{constantF}
\ee
is a solution for $a>0,c\to \infty$ and for $a<0, c=0$, which means
\begin{equation}
B \propto r^2.
\label{brsq}
\end{equation}
In fact, as we shall see, Eq. (\ref{brsq}) must hold in general. The equation of motion $B R_{tt}^f+C R^f_
{\theta\theta}=0$ takes the form \cite{Isham:1977rj}
\be
\label{gordy}
BC''-CB''+2\Delta+(CB'-BC'){\Delta'\over 2\Delta}=0.
\ee
From (\ref{casetypeII}) and (\ref{case2}), we have
\be
A={BK\over 3B-r^2}, \quad C={BJ\over 3B-r^2},\label{aacc}
\ee
and hence
\be
\Delta={B^2 \over (3B-r^2)^2}.\label{dd}
\ee
Now, Eqs. (\ref{aacc}) and (\ref{dd}) can be used in (\ref{gordy}) in order to eliminate $\Delta$ and $C$ in terms of $B$ and its derivatives (as well as the known function $J$ and its derivatives). The derivatives of $B$ can be eliminated from (\ref{eqdiffB}), and with this Eq. (\ref{gordy}) becomes an algebraic equation relating $B$ and $r$. Substituting $B=r^2F^2$, and then eliminating $r$ from Eq. (\ref{tess}), we find an algebraic equation involving only $F$ and the integration constants $a$ and $c$. It turns out that this algebraic equation does {\em not} vanish identically. Indeed, the first terms in an expansion in powers of $F$ are given by
\ba
&&B R_{tt}^f+C R^f_
{\theta\theta}=\frac{J(r)F(r)}{(3F(r)^2-1)^4 r}
\Big(\Big(\frac{\sqrt{12+a^2}-a}{\sqrt{12+a^2}+a}\Big)
^{\frac{3a}{\sqrt{12+a^2}}}c^{-3}\Lambda_g\nonumber \\
&+&\Big\{2a\Big(\frac{\sqrt{12+a^2}-a}{\sqrt{12+a^2}+a}\Big)
^{\frac{a}{\sqrt{12+a^2}}}c^{-1}+
9a\Big(\frac{\sqrt{12+a^2}-a}{\sqrt{12+a^2}+a}\Big)
^{\frac{3a}{\sqrt{12+a^2}}
}c^{-3}\Lambda_g-6M\Big\}F+O(F^2)\Big),\nonumber
\ea
where we have used $J=1-2M/r+\Lambda_g r^2/3$. For the zeroth and first order to cancel identically, one needs
\be
\Lambda_g=0, \quad M=\frac{a c^{-1}}{3}
\Big(\frac{\sqrt{12+a^2}-a}{\sqrt{12+a^2}+a}\Big)
^{\frac{a}{\sqrt{12+a^2}}},
\ee
but then going to the next order in $F$ the expression (\ref{gordy}) does not cancel for any value of $a$. Thus, $F$ is fixed to be a constant whose value is determined by (\ref{gordy}). From this (\ref{brsq}) follows.\footnote{Provided, of course, that the algebraic equation has any solution at all. Otherwise there simply aren't any solutions under the assumption (\ref{case2}). Note, in particular, from (\ref{tess}) and the subsequent discussion, that the constancy of $F$ can only be achieved for very special values of the integration constants, but these turn out to be the only relevant ones.} Now, it is easy to show that whenever $B\propto r^2$ both metrics must be proportional to each other. Indeed, it follows from Eq. (\ref{eqdiffB}) that $\Delta=AC=const.$ and $B=(a^2\Delta) r^2$. Also, using $JK=1$ and (\ref{casetypeII}) we have $A=\Delta^{1/2} K$ and $C=\Delta^{1/2} J$. On the other hand, for constant $\Delta$, Eq. (\ref{gordy}) reads$$
B C''-C B'' +2\Delta=0.
$$
Using $B=(a^2\Delta) r^2$, $C=\Delta^{1/2} J$ and $J=1-2M/r+\Lambda_g r^2/3$, where $M$ and $\Lambda_g$ are constants, it follows immediately that $a^2=\Delta^{-1/2}$, which implies $B=\Delta^{1/2} r^2$. It is then clear that $f_{\mu\nu}=\gamma g_{\mu\nu}$, where $\gamma=\Delta^{1/2}$ is a constant, as we intended to show.

%Aside from these special solutions, it is interesting to %discuss the asymptotic behaviour of $F$ as a function of %$r$ in more generic cases. For finite $c$ and $F \ll %\sqrt{12+a^2}-|a|$ the solution is $F\propto r^{-1}$, which %means that $B(r)$ will approach a constant for large $r\gg %[c (\sqrt{12+a^2}-|a|)]^{-1}$.{\bf I am not sure the %following comment will be useful or not: For $F \gg %\sqrt{12+a^2}+|a|$, we also have $F\propto r^{-1}$, and %$B(r)$ approaches a constant for small $r$. Note, however, %that the quantity in parenthesis in (\ref{tess}) becomes %negative in this limit, and so $r$ would be complex if $F$ %is real. This might be relevant if we are interested in %(\ref{tess}) as an analytic function relating $r$ and $F$, %regardless of whether this function becomes complex in some %domain}.
%Finally, for $a<0$ and finite $c$, $F$ tends to the constant
%(\ref{constantF}) as $r\to 0$, which means that $B\propto %r^2$ in this limit.

Next, let us consider the case (\ref{case1}). Here, we can use (\ref{casetypeII}) and (\ref{eqdiffB}) to
obtain \be \label{finaleqdiff} \frac{B^{\prime 2}}{B^3}\propto
\frac{1}{r^4}, \ee
and equation (\ref{finaleqdiff}) yields \be B=\frac{\gamma
r^2}{(1+\alpha r)^2}. \ee Since we have assumed that $g$ satisfies
Einstein's equations with a cosmological constant, Eq.
(\ref{Einstein}), $T_{\mu\nu}^g$
should be proportional to $g_{\mu\nu}$ with a constant
proportionality factor. This is achieved only for $\alpha=0$ which
means $C=\gamma J$. This means that both metrics will be
proportional, with \be f_{\mu\nu}=\gamma g_{\mu\nu}. \ee This completes our proof.

As discussed in the text, the remaining equations of motion determine
the constant $\gamma$ in terms of the parameters in the
Lagrangian.

%%%%%%%%%%%%%%%%%%%%%%%%%%%%%%%%%%%%%%%%%%%%%%%%%%%%%%%%%%%%%%
%%%%%%%%%%%%%%%%%%%%%%%%%%%%%%%%%%%%%%%%%%%%%%%%%%%%%%%%%%%%%%


\begin{thebibliography}{99}
%%%%%%%%%%%%%%%%%%%%%%%%%%%%%%%%%%%%%%%%%%%%%%%%%%%%%%%%%%%%%%
%%%%%%%%%%%%%%%%%%%%%%%%%%%%%%%%%%%%%%%%%%%%%%%%%%%%%%%%%%%%%%

%\cite{Dvali:2000hr}
\bibitem{Dvali:2000hr}
G.~R.~Dvali, G.~Gabadadze and M.~Porrati,
%``4D gravity on a brane in 5D Minkowski space,''
Phys.\ Lett.\ B {\bf 485} (2000) 208
[arXiv:hep-th/0005016].
%%CITATION = HEP-TH 0005016;%%

%\cite{Damour:2002ws}
\bibitem{Damour:2002ws}
T.~Damour and I.~I.~Kogan,
%``Effective Lagrangians and universality classes of nonlinear bigravity,''
Phys.\ Rev.\ D {\bf 66} (2002) 104024
[arXiv:hep-th/0206042].
%%CITATION = HEP-TH 0206042;%%
%\cite{Damour:2002wu}


%\cite{Arkani-Hamed:2003uy}
\bibitem{Arkani-Hamed:2003uy}
N.~Arkani-Hamed, H.~C.~Cheng, M.~A.~Luty and S.~Mukohyama,
%``Ghost condensation and a consistent infrared modification of gravity,''
JHEP {\bf 0405}, 074 (2004)
[arXiv:hep-th/0312099].
%%CITATION = HEP-TH 0312099;%

%\cite{Rubakov:2004eb}
\bibitem{Rubakov:2004eb}
 V.~A.~Rubakov,
  %``Lorentz-violating graviton masses: Getting around ghosts, low strong
  %coupling scale and VDVZ discontinuity,''
  arXiv:hep-th/0407104.
  %%CITATION = HEP-TH 0407104;%%

%\cite{Dubovsky:2004sg}
\bibitem{Dubovsky:2004sg}
S.~L.~Dubovsky,
%``Phases of massive gravity,''
JHEP {\bf 0410}, 076 (2004)
[arXiv:hep-th/0409124].
%%CITATION = HEP-TH 0409124;%%

%\cite{ACCDGP}
\bibitem{ACCDGP}
C.~Deffayet,
%``Cosmology on a brane in Minkowski bulk,''
Phys.\ Lett.\ B {\bf 502} (2001) 199
[arXiv:hep-th/0010186].
%%CITATION = HEP-TH 0010186;%%.
C.~Deffayet, G.~R.~Dvali and G.~Gabadadze,
%``Accelerated universe from gravity leaking to extra dimensions,''
Phys.\ Rev.\ D {\bf 65} (2002) 044023
[arXiv:astro-ph/0105068].
%%CITATION = ASTRO-PH 0105068;%%

\bibitem{Damour:2002wu}
T.~Damour, I.~I.~Kogan and A.~Papazoglou,
%``Non-linear bigravity and cosmic acceleration,''
Phys.\ Rev.\ D {\bf 66}, 104025 (2002)
[arXiv:hep-th/0206044].
%%CITATION = HEP-TH 0206044;%%
%\cite{Damour:2002gp}

%\cite{Dubovsky:2005dw}
\bibitem{Dubovsky:2005dw}
  S.~L.~Dubovsky, P.~G.~Tinyakov and I.~I.~Tkachev,
  %``Cosmological attractors in massive gravity,''
  Phys.\ Rev.\ D {\bf 72} (2005) 084011
  [arXiv:hep-th/0504067].
  %%CITATION = HEP-TH 0504067;%%
  %%Cited 4 times in SPIRES-HEP

%\cite{Fierz:1939ix}
\bibitem{Fierz:1939ix}M. Fierz, Helv. Phys. Acta 22 (1939) 3;
M.~Fierz and W.~Pauli,
%``On Relativistic Wave Equations For Particles Of Arbitrary Spin In An
%Electromagnetic Field,''
Proc.\ Roy.\ Soc.\ Lond.\ A {\bf 173}, 211 (1939).
%%CITATION = PRSLA,A173,211;%%

%\cite{Isham:gm}
\bibitem{Isham:gm}
C.~J.~Isham, A.~Salam and J.~Strathdee,
%``F-Dominance Of Gravity,''
Phys.\ Rev.\ D {\bf 3} (1971) 867.

\bibitem{DECONS}
N.~Arkani-Hamed, A.~G.~Cohen and H.~Georgi,
%``(De)constructing dimensions,''
Phys.\ Rev.\ Lett.\  {\bf 86} (2001) 4757
[arXiv:hep-th/0104005].
%%CITATION = HEP-TH 0104005;%%
%\cite{Hill:2000mu}
%\bibitem{Hill:2000mu}
C.~T.~Hill, S.~Pokorski and J.~Wang,
%``Gauge invariant effective Lagrangian for Kaluza-Klein modes,''
Phys.\ Rev.\ D {\bf 64} (2001) 105005
[arXiv:hep-th/0104035].
%%CITATION = HEP-TH 0104035;%%

\bibitem{GRAVDECONS}
N.~Arkani-Hamed, H.~Georgi and M.~D.~Schwartz,
%``Effective field theory for massive gravitons and gravity in theory  space,''
Annals Phys.\  {\bf 305} (2003) 96
[arXiv:hep-th/0210184].
%\cite{Arkani-Hamed:2003vb}
N.~Arkani-Hamed and M.~D.~Schwartz,
%``Discrete gravitational dimensions,''
Phys.\ Rev.\ D {\bf 69}, 104001 (2004)
[arXiv:hep-th/0302110].
%%CITATION = HEP-TH 0302110;%%
%\cite{Schwartz:2003vj}
M.~D.~Schwartz,
%``Constructing gravitational dimensions,''
Phys.\ Rev.\ D {\bf 68}, 024029 (2003)
[arXiv:hep-th/0303114].
%%CITATION = HEP-TH 0303114;%%
C.~Deffayet and J.~Mourad,
%``Multigravity from a discrete extra dimension,''
Phys.\ Lett.\ B {\bf 589} (2004) 48
[arXiv:hep-th/0311124].
%%CITATION = HEP-TH 0311124;%%



\bibitem{vDVZ}
%\cite{vanDam:1970vg}
%\bibitem{vanDam:1970vg}
H.~van Dam and M.~J.~Veltman,
%``Massive And Massless Yang-Mills And Gravitational Fields,''
Nucl.\ Phys.\ B {\bf 22} (1970) 397.
%%CITATION = NUPHA,B22,397;%%
%\bibitem{ZAK}
V.I.Zakharov, JETP Lett {\bf 12}, 312 (1970).
%\cite{Iwasaki:uz}
%\bibitem{Iwasaki:uz}
Y.~Iwasaki,
%``Consistency Condition For Propagators,''
Phys.\ Rev.\ D {\bf 2} (1970) 2255.
%%CITATION = PHRVA,D2,2255;%%

%\cite{Vainshtein:1972sx}
\bibitem{Vainshtein:1972sx}
A.~I.~Vainshtein,
%``To The Problem Of Nonvanishing Gravitation Mass,''
Phys.\ Lett.\ B {\bf 39} (1972) 393.
%%CITATION = PHLTA,B39,393;%%

\bibitem{CUREVDVZ}
%\cite{Vainshtein:sx}
%\bibitem{Vainshtein:sx}
%A.~I.~Vainshtein,
%``To The Problem Of Nonvanishing Gravitation Mass,''
%Phys.\ Lett.\ B {\bf 39} (1972) 393.
%%CITATION = PHLTA,B39,393;%%
%\bibitem{Deffayet:2001uk}
%\cite{Dvali:2000hr}
%\bibitem{Dvali:2000hr}
G.~R.~Dvali, G.~Gabadadze and M.~Porrati,
%``4D gravity on a brane in 5D Minkowski space,''
Phys.\ Lett.\ B {\bf 485}, 208 (2000)
%[arXiv:hep-th/0005016].
%%CITATION = HEP-TH 0005016;%%
C.~Deffayet, G.~R.~Dvali, G.~Gabadadze and A.~I.~Vainshtein,
%``Nonperturbative continuity in graviton mass versus perturbative  discontinuity,''
Phys.\ Rev.\ D {\bf 65} (2002) 044026
[arXiv:hep-th/0106001].
%%CITATION = HEP-TH 0106001;%%
%\bibitem{ARTHUR}
A.~Lue,
%``Cosmic strings in a brane world theory with metastable gravitons,''
Phys.\ Rev.\ D {\bf 66} (2002) 043509
[arXiv:hep-th/0111168].
%%CITATION = HEP-TH 0111168;%%
%\cite{Gruzinov:2001hp}
%\bibitem{ANDRE}
A.~Gruzinov,
%``On the graviton mass,''
New\ Astron. \ {\bf 10} (2005) 311 [arXiv:astro-ph/0112246].
%%CITATION = ASTRO-PH 0112246;%%
%\cite{Porrati:2002cp}
%\bibitem{MASSIMO}
M.~Porrati,
%``Fully covariant van Dam-Veltman-Zakharov discontinuity, and absence  thereof,''
Phys.\ Lett.\ B {\bf 534} (2002) 209
[arXiv:hep-th/0203014].
%%CITATION = HEP-TH 0203014;%%
%\cite{Tanaka:2003zb}
%\bibitem{Tanaka:2003zb}
  T.~Tanaka,
  %``Weak gravity in DGP braneworld model,''
  Phys.\ Rev.\ D {\bf 69} (2004) 024001
  [arXiv:gr-qc/0305031].
  %%CITATION = GR-QC 0305031;%%
  %%Cited 22 times in SPIRES-HEP
%\cite{Gabadadze:2004iy}
%\bibitem{Gabadadze:2004iy}
  G.~Gabadadze and A.~Iglesias,
  %``Schwarzschild solution in brane induced gravity,''
  Phys.\ Rev.\ D {\bf 72} (2005) 084024
  [arXiv:hep-th/0407049].
  %%CITATION = HEP-TH 0407049;%%


\bibitem{Damour:2002gp}
T.~Damour, I.~I.~Kogan and A.~Papazoglou,
%``Spherically symmetric spacetimes in massive gravity,''
Phys.\ Rev.\ D {\bf 67}, 064009 (2003)
[arXiv:hep-th/0212155].
%%CITATION = HEP-TH 0212155;%%




%\cite{Boulware:my}
\bibitem{Boulware:my}
D.~G.~Boulware and S.~Deser,
%``Can Gravitation Have A Finite Range?,''
Phys.\ Rev.\ D {\bf 6} (1972) 3368.
%%CITATION = PHRVA,D6,3368;%%

%\cite{Deffayet:2005ys}
\bibitem{Deffayet:2005ys}
  C.~Deffayet and J.~W.~Rombouts,
  %``Ghosts, strong coupling and accidental symmetries in massive gravity,''
  Phys.\ Rev.\ D {\bf 72} (2005) 044003
  [arXiv:gr-qc/0505134].
  %%CITATION = GR-QC 0505134;%%
  %%Cited 5 times in SPIRES-HEP

%\cite{Creminelli:2005qk}
\bibitem{Creminelli:2005qk}
  P.~Creminelli, A.~Nicolis, M.~Papucci and E.~Trincherini,
  %``Ghosts in massive gravity,''
  JHEP {\bf 0509} (2005) 003
  [arXiv:hep-th/0505147].
  %%CITATION = HEP-TH 0505147;%%
  %%Cited 5 times in SPIRES-HEP


%\cite{Salam:1976as}
\bibitem{Salam:1976as}
A.~Salam and J.~Strathdee,
%``A Class Of Solutions For The Strong Gravity Equations,''
Phys.\ Rev.\ D {\bf 16}, 2668 (1977).


%\cite{Isham:1977rj}
\bibitem{Isham:1977rj}
C.~J.~Isham and D.~Storey,
%``Exact Spherically Symmetric Classical Solutions For The F-G Theory Of Gravity,''
Phys.\ Rev.\ D {\bf 18}, 1047 (1978).

%%CITATION = HEP-TH 0311125;%%
\bibitem{arache}
C. Aragone and J. Chela-Flores, Nuovo Cimento {\bf 10A} 818
(1972); J. Chela-Flores, Int. J. Theor. Phys. {\bf 10}, 103 (1974)

%\cite{vanderBij:1981uw}
\bibitem{vanderBij}
J.~J.~van der Bij, H.~van Dam and Y.~J.~Ng,% ``Theory Of Gravity
%And The Cosmological Term: The Little Group Viewpoint,'' Physica
{\bf 116A}, 307 (1982).
%%CITATION = PHYSA,116A,307;%%

\bibitem{Zee:1985}
A.~Zee, in {\it High Energy Physics: Proceedings of the 20th Annual Orbis Scientiae}, 1983 edited by S.L. Mintz and A. Perlmutter (Plenum, New York).
%\cite{Buchmuller:1988wx}
%\bibitem{Buchmuller:1988wx}
W.~Buchmuller and N.~Dragon,
%``Einstein Gravity From Restricted Coordinate Invariance,''
Phys.\ Lett.\ B {\bf 207}, 292 (1988).
%%CITATION = PHLTA,B207,292;%%
%\cite{Buchmuller:1988yn}
%\bibitem{Buchmuller:1988yn}
W.~Buchmuller and N.~Dragon,
%``Gauge Fixing And The Cosmological Constant,''
Phys.\ Lett.\ B {\bf 223}, 313 (1989).
%%CITATION = PHLTA,B223,313;%%

%\cite{Weinberg:1988cp}
\bibitem{Weinberg:1988cp}
S.~Weinberg,
%``The Cosmological Constant Problem,''
Rev.\ Mod.\ Phys.\  {\bf 61}, 1 (1989).
%%CITATION = RMPHA,61,1;%%

%\cite{Gabadadze:2003jq}
\bibitem{Gabadadze:2003jq}
  G.~Gabadadze and A.~Gruzinov,
  %``Graviton mass or cosmological constant?,''
  Phys.\ Rev.\ D {\bf 72} (2005) 124007
  [arXiv:hep-th/0312074].
  %%CITATION = HEP-TH 0312074;%%
  %%Cited 15 times in SPIRES-HEP



\bibitem{BDG}
D.~Blas, C~.Deffayet, J.~Garriga, {\it in preparation}.

%\cite{Deffayet:2004ws}
\bibitem{Deffayet:2004ws}
C.~Deffayet and J.~Mourad,
%``Solutions of multigravity theories and discretized brane worlds,''
Class.\ Quant.\ Grav.\  {\bf 21}, 1833 (2004)
[arXiv:hep-th/0311125].


%%%%%%%%%%%%%%%%%%%%%%%%%%%%%%%%%%%%%%%%%%%%%%%%%%%%%%%%%%%%
% ordered references first appearing in section 3.1
%%%%%%%%%%%%%%%%%%%%%%%%%%%%%%%%%%%%%%%%%%%%%%%%%%%%%%%%%%%

%\cite{Gibbons:1977mu}
\bibitem{Gibbons:1977mu}
  G.~W.~Gibbons and S.~W.~Hawking,
  %``Cosmological Event Horizons, Thermodynamics, And Particle Creation,''
  Phys.\ Rev.\ D {\bf 15} (1977) 2738.

%%%%%%%%%%%%%%%%%%%%%%%%%%%%%%%%%%%%%%%%%%%%%%%%%%%%%%%%%%%%%%%
%%%% above this line, references have been ordered
%%%%%%%%%%%%%%%%%%%%%%%%%%%%%%%%%%%%%%%%%%%%%%%%%%%%%%%%%%%%%%

\end{thebibliography}
\end{document}